\documentclass{aims}
\usepackage{amsmath}
  \usepackage{paralist}
\usepackage{listings}
\usepackage{listings,xcolor,courier}
\usepackage{hhline}
\usepackage{array}
  \usepackage{graphicx} 
  \usepackage{epsfig} 
 \usepackage[colorlinks=true]{hyperref}
\hypersetup{urlcolor=blue, citecolor=red}

\def\currentvolume{X}

    \def\ppages{X--XX}
\theoremstyle{definition}

\title[Calculating Ellipse Overlap Areas]
      {Calculating Ellipse Overlap Areas}

\author[Gary B. Hughes and Mohcine Chraibi]{}


 \email{gbhughes@calpoly.edu}
 \email{ m.chraibi@fz-juelich.de}

\begin{document}
\maketitle

\centerline{\scshape Gary B. Hughes }
\medskip
{\footnotesize
 \centerline{ California Polytechnic State University}
   \centerline{ Statistics Department}
   \centerline{ San Luis Obispo, CA 93407-0405, USA}
} 

\medskip

\centerline{\scshape Mohcine Chraibi}
\medskip
{\footnotesize
 \centerline{ J\"ulich Supercomputing Centre}
   \centerline{ Forschungszentrum J\"ulich GmbH}
   \centerline{ D-52425 J\"ulich, Germany}
}

\bigskip


\begin{abstract}
We present a general algorithm for finding the overlap area between two ellipses.  The algorithm is based on finding a segment area (the area between an ellipse and a secant line) given two points on the ellipse.  The Gauss-Green formula is used to determine the ellipse sector area between two points, and a triangular area is added or subtracted to give the segment area.  For two ellipses, overlap area is calculated by adding the areas of appropriate sectors and polygons.  Intersection points for two general ellipses are found using Ferrari's quartic formula to solve the polynomial that results from combining the two ellipse equations.  All cases for the number of intersection points (0, 1, 2, 3, 4) are handled.  The algorithm is implemented in c-code, and has been tested with a range of input ellipses.  The code is efficient enough for use in simulations that require many overlap area calculations.
\end{abstract}

\section{Introduction}




Ellipses are useful in many applied scenarios, and in widely disparate fields.  In our research, which happens to be in two very different areas, we have encountered a common need for efficiently calculating the overlap area between two ellipses.

In one case, the design for a solar calibrator on-board an orbiting satellite required an efficient algorithm for ellipse overlap area.  Imaging systems aboard satellites rely on semi-conductor detectors whose performance changes over time due to many factors.  To produce consistent data, some means of calibrating the detectors is required; see, e.g., \cite{Kent}.  Some systems use the sun as a light source for calibration.  In a typical solar calibrator, incident sunlight passes through an attenuator grating and impinges on a diffuser plate, which is oriented obliquely to the attenuator grating.  The attenuator grating is a pattern of circular openings.  When sunlight passes through the circular openings, projections of the circles onto the oblique diffuser plate become small ellipses.  The projection of the large circular entrance aperture on the oblique diffuser plate is also an ellipse.  The total incident light on the calibrator is proportional to the sum of all the areas of the smaller ellipses that are contained within the larger entrance aperture ellipse.  However, as the calibration process proceeds, the satellite is moving through its orbit, and the angle from the sun into the calibrator changes (\~{}7${}^\circ$ in 2 minutes).  The attenuator grating ellipses thus move across the entrance aperture, and some of the smaller ellipses pass in and out of the entrance aperture ellipse during calibration.  Movement of the small ellipses across the aperture creates fluctuations in the total amount of incident sunlight reaching the calibrator in the range of 0.3 to 0.5\%.  This jitter creates errors in the calibration algorithms.  In order to model the jitter, an algorithm is required for determining the overlap area of two ellipses.  Monte Carlo integration had been used; however, the method is numerically intensive because it converges very slowly, so it was not an attractive approach for modeling the calibrator due to the large number of ellipses that must be modeled.

In a more down-to-earth setting, populated places such as city streets or building corridors can become quite congested while crowds of people are moving about.  Understanding the dynamics of pedestrian movement in these scenarios can be beneficial in many ways.  Pedestrian dynamics can provide critical input to the design of buildings or city infrastructure, for example by predicting the effects of specific crowd management strategies, or the behavior of crowds utilizing emergency escape routes.  Current research in pedestrian dynamics is making steady progress toward realistic modeling of local movement; see, e.g., \cite{Chraibi2010a}.  The model presented in \cite{Chraibi2010a} is based on the concept of elliptical volume exclusion for individual pedestrians.  Each model pedestrian is surrounded by an elliptical footprint area that the model uses to anticipate obstacles and other pedestrians in or near the intended path.  The footprint area is influenced by an individuals' velocity; for example, the exclusion area in front of a fast-moving pedestrian is elongated when compared to a slower-moving individual, since a pedestrian is generally thinking a few steps ahead.  As pedestrians travel through a confined space, their collective exclusion areas become denser, and the areas will eventually begin to overlap.  A force-based model will produce a repulsive force between overlapping exclusion areas, causing the pedestrians to slow down or change course when the exclusion force becomes large.  Implementing the force-based model with elliptical exclusion areas in a simulation requires calculating the overlap area between many different ellipses in the most general orientations.  The ellipse area overlap algorithm must also be efficient, so as not to bog down the simulation.

Simulations for both the satellite solar calibrator and force-based pedestrian dynamic model require efficient calculation of the overlap area between two ellipses.  In this paper, we provide an algorithm that has served well for both applications.  The core component of the overlap area algorithm is based on determining the area of an \textit{ellipse segment}, which is the area between a secant line and the ellipse boundary.  The segment algorithm forms the basis of an application for calculating the overlap area between two general ellipses.

\section{Ellipse area, sector area and segment area}
\subsection{Ellipse Area}

Consider an ellipse that is centered at the origin, with its axes aligned to the coordinate axes.  If the semi-axis length along the \textit{x}-axis is \textit{A}, and the semi-axis length along the \textit{y}-axis is \textit{B}, then the ellipse is defined by a locus of points that satisfy the implicit polynomial equation
\begin{equation}
\frac{x^2}{A^2}+\frac{y^2}{B^2}=1
\end{equation}
The same ellipse can be defined parametrically by:
\begin{equation}
\left. \begin{array}{c}
x=A\cdot \cos(t) \\ 
y=B\cdot \sin(t) \end{array}
\right\}\ \ 0\le t\le 2\pi 
\end{equation}
The area of such an ellipse can be found using the parameterized form with the Gauss-Green formula:

\begin{equation}\label{Multi}
 \begin{split}
\text{Area}=&\frac{1}{2}\int^B_A{[x(t)\cdot y^{\prime}(t)-y(t)\cdot x{^\prime}(t)]dt}\\
=&\frac{1}{2}\int^{2\pi}_0{A\cdot \cos(t)\cdot B\cdot \cos(t)-B\cdot \sin(t)\cdot (-A)\cdot \sin(t)]dt} \\
=&\frac{A\cdot B}{2}\int^{2\pi }_0{\cos^2(t)+\sin^2(t)]dt}=\frac{A\cdot B}{2}\int^{2\pi }_0{dt}\\
=&\pi \cdot A\cdot B
 \end{split}
\end{equation}

\subsection{Ellipse Sector Areas}

\begin{figure}[htp]
\begin{center}
  \includegraphics[width=2in]{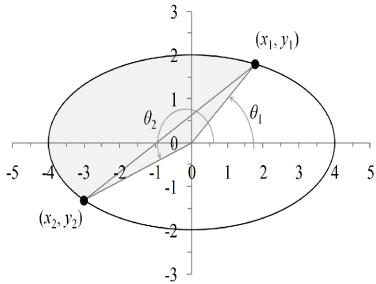}\\
  \caption{The area of an \textit{ellipse sector} between two points on the ellipse is the area swept out by a vector from the origin to the first point as the vector travels along the ellipse in a counter-clockwise direction to the second point.  The area of an ellipse sector can be determined with the Gauss-Green formula, using the parametric angles $\theta_{1}$ and $\theta_{2}$.}
\label{fig1}
  \end{center}
\end{figure}

We define the \textit{ellipse sector} between two points (${x}_{1}$, ${y}_{1}$) and (${x}_{2}$, ${y}_{2}$) on the ellipse as the area that is swept out by a vector from the origin to the ellipse, beginning at (${x}_{1}$, ${y}_{1}$), as the vector travels along the ellipse in a counter-clockwise direction from (${x}_{1}$, ${y}_{1}$) to (${x}_{2}$, ${y}_{2}$).  An example is shown in Fig.~\ref{fig1}. The Gauss-Green formula can also be used to determine the area of such an ellipse sector.

\begin{equation}\label{Multi}
 \begin{split}
    \text{Sector Area}=& \frac{A\cdot B}{2}\int^{\theta_2}_{\theta_1}{dt}\\
    =& \frac{(\theta_2-\theta_1)\cdot A\cdot B}{2}
 \end{split}
\end{equation}

The parametric angle $\theta $ that is formed between the $x$-axis and a point ($x$, $y$) on the ellipse is found from the ellipse parameterizations:
\begin{align*}
x=&A\cdot \cos(\theta)\ \ \Longrightarrow \theta =\cos^{-1}(x/A)\\
y=&B\cdot \sin(\theta)\ \ \Longrightarrow \theta =\sin^{-1}(y/B)\ 
\end{align*}

For a circle ($A = B$ in the ellipse implicit polynomial form), the parametric angle corresponds to the geometric (visual) angle that a line from the origin to the point ($x$, $y$) makes with the $x$-axis.  However, the same cannot be said for an ellipse; that is, the geometric (visual) angle is \textit{not} the same as the parametric angle used in the area calculation.  For example, consider the ellipse in Fig.~\ref{fig1}; the implicit polynomial form is
\begin{equation}
\frac{x^2}{4^2}+\frac{y^2}{2^2}=1
\end{equation}
Suppose the point ($x_1$, $y_1$) is at $\left({4}/{\sqrt{5}},{4}/{\sqrt{5}}\right)$.  The point is on the ellipse, since
\begin{equation*}
\frac{{\left({4}/{\sqrt{5}}\right)}^2}{4^2}+\frac{{\left({4}/{\sqrt{5}}\right)}^2}{2^2}=\frac{{4^2}/{5}}{4^2}+\frac{{4^2}/{5}}{2^2}=\frac{1}{5}+\frac{4}{5}=1
\end{equation*}

A line segment from the origin to $\left({4}/{\sqrt{5}},{4}/{\sqrt{5}}\right)$ forms an angle with the $x$-axis of $\pi $/4 ($\approx$0.7485398).  However, the ellipse parametric angle to the same point is:
\begin{equation*}
\theta ={{\cos }^{-1} \left(\frac{{4}/{\sqrt{5}}}{4}\right)\ }={{\cos }^{-1} \left(\frac{1}{\sqrt{5}}\right)\ }\approx 1.10715
\end{equation*}
The same angle can also be found from the parametric equation for $y$:
\begin{equation*}
\theta ={\sin^{-1} \left(\frac{4/\sqrt{5}}{2}\right)\ }={\sin^{-1} \left(\frac{2}{\sqrt{5}}\right)\ }\approx 1.10715
\end{equation*}
The angle found by using the parametric equations does not match the geometric angle to the point that defines the angle.

When determining the parametric angle for a given point ($x$, $y$) on the ellipse, the angle must be chosen in the proper quadrant, based on the signs of $x$ and $y$.  For the ellipse in Fig.~\ref{fig1}, suppose the point ($x_{2}$, $y_{2}$) is at $\left(-3,-{\sqrt{7}}/{2}\right)$.  The parametric angle that is determined from the equation for $x$ is:
\begin{equation*}
\theta ={\cos^{-1} \left(\frac{-3}{4}\right)\ }\approx 2.41886
\end{equation*}

The parametric angle that is determined from the equation for $y$ is:

\begin{equation*}
\theta ={{\sin }^{-1} \left(\frac{-{\sqrt{7}}/{2}}{2}\right)\ }={{\sin }^{-1} \left(\frac{-\sqrt{7}}{4}\right)\ }\approx -.722734
\end{equation*}

The apparent discrepancy is resolved by recalling that inverse trigonometric functions are usually implemented to return a `principal value' that is within a conventional range.  The typical (principal-valued) $\theta  = \arccos(x)$ function returns angles in the range 0 = $\theta $ = $\pi $, and the typical (principal-valued) $\theta  = \arcsin (x)$ function returns angles in the range --$\pi /2 = \theta  = \pi/2$.  When the principal-valued inverse trigonometric functions return angles in the typical ranges, the ellipse parametric angles, defined to be from the $x$-axis, with positive angles in the counter-clockwise direction, can be found with the relations in Table \ref{tab1}.

\begin{table}[]
\begin{tabular}{|p{2.0in}|p{1.7in}|} \hline 
\textbf{Quadrant II} ($x< 0\; \text{and}\; y \geq 0$)\newline $\theta =\arccos(x/A)$\newline $=\pi -\arcsin(y/B)$ & \textbf{Quadrant I} ($x \geq 0\; \text{and}\; y \geq 0)$\newline $\theta =\arccos(x/A)$\newline $=\arcsin(y/B)$ \\ \hline 
\textbf{Quadrant III} ($x< 0\; \text{and}\; y < 0$)\newline $\theta =2\pi -\arccos(x/A)$\newline $=\pi -\arcsin(y/B)$ & \textbf{Quadrant IV} ($x \geq 0\; \;y < 0$)\newline $\theta =2\pi -\arccos(x/A)$\newline $=2\pi +\arcsin(y/B)$ \\ \hline 
\end{tabular}
\label{tab1}
\caption{Relations for finding the parametric angle that corresponds to a given point (\textit{x}, \textit{y}) on the ellipse \textit{x}${}^{2}$/\textit{A}${}^{2}$ + \textit{y}${}^{2}$/\textit{B}${}^{2}$ = 1.  The parametric angle is formed between the positive \textit{x}-axis and a line drawn from the origin to the given point, with counterclockwise being positive.  For the standard (principal-valued) inverse trigonometric functions, the resulting angle will be in the range 0 $\leq\theta <2\pi $ for any point on the ellipse.}
\end{table}

The point at $\left(-3,-{\sqrt{7}}/{2}\right)$ on the ellipse of Fig.~\ref{fig1} is in Quadrant III.  Using the relations in Table~\ref{tab1}, the parametric angle that is determined from the equation for $x$ is:
\[\theta =2\pi -\arccos(\frac{-3}{4})\approx 3.86433\] 
The parametric angle that is determined from the equation for $y$ is:

\begin{equation*}
\theta =\pi -\arcsin(\frac{ -\sqrt{7}/2 }{2})\approx 3.86433
\end{equation*} 

With the proper angles, the Gauss-Green formula can be used to determine the area of the sector from the point at $\left({4}/{\sqrt{5}},{4}/{\sqrt{5}}\right)$ to the point $\left(-3,-{\sqrt{7}}/{2}\right)$ in the ellipse of Fig.~\ref{fig1}.

\begin{equation}\label{Multi}
 \begin{split}
    \text{Sector Area}=&\frac{\left({\theta }_2-{\theta }_1\right)\cdot A\cdot B}{2}\\
    =& \frac{\left[\left(2\pi -\arccos\left(\frac{-3}{4}\right)\right)-{\arccos  \left(\frac{{4}/{\sqrt{5}}}{4}\right)\ }\right]\cdot 4\cdot 2}{2}\\
\approx& 11.0287
 \end{split}
\end{equation}

\begin{figure}[htp]
\begin{center}
  \includegraphics[width=2in]{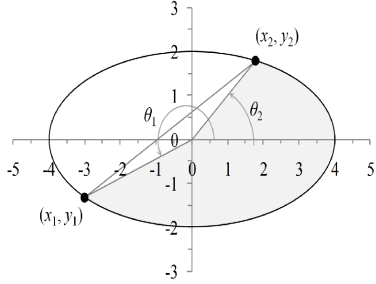}\\
  \caption{We define the ellipse sector area to be calculated from the first point ($x_1$, $y_1$) to the second point ($x_2$, $y_2$) in a counter-clockwise direction along the ellipse.}
\label{fig2}
  \end{center}
\end{figure}

The Gauss-Green formula is sensitive to the direction of integration.  For the larger goal of determining ellipse overlap areas, we define the ellipse sector area to be calculated from the first point ($x_{1}$, $y_{1}$) to the second point ($x_{2}$, $y_{2}$) in a \textit{counter-clockwise} direction along the ellipse.  For example, if the points ($x_{1}$, $x_{1}$) and ($x_{2}$, $y_{2}$) of Fig.~\ref{fig1} were to have their labels switched, then the ellipse sector defined by the new points will have an area that is complementary to that of the sector in Fig.~\ref{fig1}, as shown in Fig.~\ref{fig2}.

Switching the point labels, as shown in Fig.~\ref{fig2}, also causes the angle labels to be switched, resulting in the condition that $\theta_{1}>\theta_{2}$.  Since using the definitions in Table~\ref{tab1} will always produce an angle in the range $0 = \theta < 2\pi $ for any point on the ellipse, the first angle can be transformed by subtracting 2$\pi $ to restore the condition that $\theta_{1}<\theta_{2}$.  The sector area formula given above can then be used, with the integration angle from ($\theta_{1}$ -- $2\pi $) through\, $\theta_{2}$.  With the angle labels shown in Fig.~\ref{fig2}, the area of the sector from the point at $\left(-3,-{\sqrt{7}}/{2}\right)$ to the point at $\left({4}/{\sqrt{5}},{4}/{\sqrt{5}}\right)$ in a counter-clockwise direction is:
\begin{equation}\label{Multi}
 \begin{split}
    \text{Sector Area}=&\frac{\left(\theta_2-\left(\theta_1-2\pi \right)\right)\cdot A\cdot B}{2}\\
    =& \frac{\left[\left(2\pi -\arccos\left(\frac{-3}{4}\right)\right)-\left({\arccos  \left(\frac{{4}/{\sqrt{5}}}{4}\right)-2\pi \ }\right)\right]\cdot 4\cdot 2}{2}\\
\approx& 14.1040
 \end{split}
\end{equation}

The two sector areas shown in Fig.~\ref{fig1} and Fig.~\ref{fig2} are complementary, in that they add to the total ellipse area.  Using the angle labels as shown in Fig.~\ref{fig1} for both sector areas:
\begin{equation}\label{Multi}
 \begin{split}
    \text{Total Area}=&\frac{\left(\theta_2-\theta_1\right)\cdot A\cdot B}{2}+\frac{\left(\theta_1-\left(\theta_2-2\pi \right)\right)\cdot A\cdot B}{2}\\
    =& \frac{\left(2\pi \right)\cdot A\cdot B}{2}=\pi \cdot A\cdot B\\
    =&\pi \cdot 4\cdot 2\\
\approx& 25.1327
 \end{split}
\end{equation}

\subsection{Ellipse Segment Areas}

For the overall goal of determining overlap areas between ellipses and other curves, a useful measure is the area of what we will call an \textit{ellipse} \textit{segment}.  A secant line drawn between two points on an ellipse partitions the ellipse area into two fractions, as shown in Fig.~\ref{fig1} and Fig.~\ref{fig2}.  We define the ellipse segment as the area confined by the secant line and the portion of the ellipse from the first point ($x_{1}$, $y_{1}$) to the second point ($x_{2}$, $y_{2}$) \textit{traversed in a counter-clockwise direction}.  The segment's complement is the second of the two areas that are demarcated by the secant line.  For the ellipse of Fig.~\ref{fig1}, the area of the segment defined by the secant line through the points ($x_{1}$, $y_{1}$) and ($x_{2}$, $y_{2}$) is the area of the sector \textit{minus} the area of the triangle defined by the two points and the ellipse center.  To find the area of the triangle, suppose that the coordinates for the vertices of are known, e.g., as ($x_{1}$, $y_{1}$), ($x_{2}$, $y_{2}$) and ($x_{3}$, $y_{3}$).  Then the triangle area can be found by:
\begin{equation}
\begin{split}
\text{Triangle Area}=&\frac{1}{2}\cdot \left|det\left( \begin{array}{ccc}
x_1 & x_2 & x_3 \\ 
y_1 & y_2 & y_3 \\ 
1 & 1 & 1 \end{array}
\right)\right|\\
=&\frac{1}{2}\cdot \left|x_1\cdot \left(y_2-y_3\right)-x_2\cdot \left(y_1-y_3\right)+x_3\cdot \left(y_1-y_2\right)\right| 
\end{split}
\end{equation}
In the case where one vertex, say ($x_{3}$, $y_{3}$), is at the origin, then the area formula for the triangle can be simplified to:
\begin{equation}
\text{Triangle Area}=\frac{1}{2}\cdot \left|x_1\cdot y_2-x_2\cdot y_1\right|
\end{equation}

For the case depicted in Fig.~\ref{fig1}, subtracting the triangle area from the area of the ellipse sector area gives the area between the secant line and the ellipse, i.e., the area of the ellipse segment counter-clockwise from ($x_{1}$, $y_{1}$) to ($x_{2}$, $y_{2}$):
\begin{equation}
\text{Segment\ Area}=\frac{\left({\theta }_2-{\theta }_1\right)\cdot A\cdot B}{2}-\frac{1}{2}\cdot \left|x_1\cdot y_2-x_2\cdot y_1\right|
\end{equation}
For the ellipse of Fig.~\ref{fig1}, with the points at $\left({4}/{\sqrt{5}},{4}/{\sqrt{5}}\right)$ and $\left(-3,-{\sqrt{7}}/{2}\right)$, the area of the segment defined by the secant line is:

\begin{equation*}
\begin{split}
&\frac{\left[\left(2\pi -\arccos\left(\frac{-3}{4}\right)\right)-{\arccos  \left(\frac{{4}/{\sqrt{5}}}{4}\right)\ }\right]\cdot 4\cdot 2}{2}-\frac{1}{2}\cdot \left|\frac{4}{\sqrt{5}}\cdot \frac{-\sqrt{7}}{2}-\frac{4}{\sqrt{5}}\cdot -3\right|\\ 
&\approx 9.52865
\end{split}
\end{equation*}

For the ellipse of Fig.~\ref{fig2}, the area of the segment shown is the sector area \textit{plus} the area of the triangle.

\begin{equation}
\text{Segment Area}=\frac{\left({\theta }_2-\left({\theta }_1-2\pi \right)\right)\cdot A\cdot B}{2}+\frac{1}{2}\cdot \left|x_1\cdot y_2-x_2\cdot y_1\right|
\end{equation}

With the points at $\left(-3,-{\sqrt{7}}/{2}\right)$ and $\left({4}/{\sqrt{5}},{4}/{\sqrt{5}}\right)$ the area of the segment is:
\begin{equation*}
\begin{split}
&\frac{\left[\left(2\pi -\arccos\left(\frac{-3}{4}\right)\right)-\left({\arccos  \left(\frac{{4}/{\sqrt{5}}}{4}\right)-2\pi \ }\right)\right]\cdot 4\cdot 2}{2}+\frac{1}{2}\cdot \left|\frac{4}{\sqrt{5}}\cdot \frac{-\sqrt{7}}{2}-\frac{4}{\sqrt{5}}\cdot -3\right|\\
&\approx 15.60409411
\end{split}
\end{equation*}

For the case shown in Fig.~\ref{fig1} and Fig.~\ref{fig2}, the sector areas were shown to be complementary.  The segment areas are also complementary, since the triangle area is added to the sector of Fig.~\ref{fig1}, but subtracted from the sector of Fig.~\ref{fig2}.  Using the angle labels as shown in Fig.~\ref{fig1} for both sector areas:

\begin{equation}
\begin{split}
\text{Total\ Area}=&\left[\frac{\left({\theta }_2-{\theta }_1\right)\cdot A\cdot B}{2}-\frac{1}{2}\cdot \left|x_1\cdot y_2-x_2\cdot y_1\right|\right]\\
+&\left[\frac{\left({\theta }_1-\left({\theta }_2-2\pi \right)\right)\cdot A\cdot B}{2}+\frac{1}{2}\cdot \left|x_1\cdot y_2-x_2\cdot y_1\right|\right]\\
=&\pi \cdot A\cdot B=\pi \cdot 4\cdot 2\approx 25.1327
\end{split}
\end{equation}

The key difference between the cases in Fig.~\ref{fig1} and Fig.~\ref{fig2} that requires the area of the triangle to be either subtracted from, or added to, the sector area is the size of the \textit{integration angle}.  If the integration angle is less than $\pi$, then the triangle area must be subtracted from the sector area to give the segment area.  If the integration angle is greater than $\pi$, the triangle area must be added to the sector area.

\subsection{A Core Algorithm for Ellipse Segment Area}

A generalization of the cases given in Fig.~\ref{fig1} and Fig.~\ref{fig2} suggests a robust approach for determining the ellipse segment area defined by a secant line drawn between two given points on the ellipse.  The ellipse is assumed to be centered at the origin, with its axes parallel to the coordinate axes.  We define the segment area to be demarcated by the secant line and the ellipse proceeding counter-clockwise from the first given point ($x_{1}$, $y_{1}$) to the second given point ($x_{2}$, $y_{2}$).  The ELLIPSE\_SEGMENT algorithm is outlined in Table~\ref{tab2}, with pseudo-code presented in List.~\ref{code1}.  The ellipse is passed to the algorithm by specifying the semi-axes lengths, $A >0$\, and $B> 0$.  The points are passed to the algorithm as ($x_{1}$, $x_{1}$) and ($x_{2}$, $y_{2}$), which must be on the ellipse.

\begin{table}
\begin{center}
\begin{tabular}{|p{4.1in}|} \hline 
ELLIPSE\_SEGMENT Area Algorithm:\newline $1.\ \ \ \ \ \ \ \  \begin{array}{c}
{\theta }_1=\left\{ \begin{array}{c}
{\arccos  \left({x_1}/{A}\right)\ }\ \ \ \ \ \ \ \ \ ,\ \ y_1\ge 0 \\ 
2\pi -{\arccos  \left({x_1}/{A}\right)\ },\ \ y_1<0 \end{array}
\right. \\ 
{\theta }_2=\left\{ \begin{array}{c}
{\arccos  \left({x_2}/{A}\right)\ }\ \ \ \ \ \ \ \ \ ,\ \ y_2\ge 0 \\ 
2\pi -{\arccos  \left({x_2}/{A}\right)\ },\ \ y_2<0 \end{array}
\right. \end{array}
$\newline $2.\ \ \ \ \ \ \ \ {\widehat{\theta }}_1=\left\{ \begin{array}{c}
  {\theta }_1\ \ \ \ \ \ \ \ ,\ \ {{\theta }_1<\theta }_2 \\ 
{\theta }_1-2\pi ,\ \ {{\theta }_1>\theta }_2 \end{array}
\right.$\newline $3.\ \ \ \ \ \ \ \ {\rm Area}=\frac{\left({\theta }_2-{\widehat{\theta }}_1\right)\cdot A\cdot B}{2}+\frac{sign\left({\theta }_2-{\widehat{\theta }}_1-\pi \right)}{2}\cdot \left|x_1\cdot y_2-x_2\cdot y_1\right|$ \\ \hline 
where:\newline the ellipse implicit polynomial equation is\newline $\frac{x^2}{A^2}+\frac{y^2}{B^2}=1$\newline \textit{A} $>$ 0 is the semi-axis length along the \textit{x}-axis\newline \textit{B} $>$ 0 is the semi-axis length along the \textit{y}-axis\newline (\textit{x}${}_{1}$, \textit{y}${}_{1}$) is the first given point on the ellipse\newline ($x_{2}$, $y_{2}$) is the second given point on the ellipse\newline $\theta_{1}$ and $\theta_{2}$ are the parametric angles corresponding to the points\\ ($x_{1}$, $y_{1}$) and ($x_{2}$, $y_{2}$) \\ \hline 
\end{tabular}
\end{center}
\caption{An outline of the ELLIPSE\_SEGMENT area algorithm.}
\label{tab2}
\end{table}



For robustness, the algorithm should avoid divide-by-zero and inverse-trigonometric errors, so data checks should be included.  The ellipse parameters $A$\, and $B$ must be greater than zero.  A check is provided to determine whether the points are on the ellipse, to within some numerical tolerance, $\varepsilon$.  Since the points can only be checked as being on the ellipse to within some numerical tolerance, it may still be possible for the $x$-values to be slightly larger than $A$, leading to an error when calling the inverse trigonometric functions with the argument $x/A$.  In this case, the algorithm checks whether the $x$-value close to $A$ or --$A$, that is within a distance that is less than the numerical tolerance.  If the closeness condition is met, then the algorithm assumes that the calling function passed a value that is indeed on the ellipse near the point ($A$, 0) or (--$A$, 0), so the value of $x$ is nudged back to $A$ or --$A$ to avoid any error when calling the inverse trigonometric functions.  The core algorithm, including all data checks, is shown in List.~\ref{code1}.

\lstset{language=FORTRAN, 
      stringstyle=\color{black},
  keywordstyle=\color{cyan},
caption={The ELLIPSE\_SEGMENT algorithm is shown for calculating the area
of a segment defined by the secant line drawn between two given points ($x_1$, $y_1$)
and ($x_2$, $y_2$) on the ellipse $x_2/A_2 + y_2/B_2 = 1$. We define the segment area for this
algorithm to be demarcated by the secant line and the ellipse proceeding counter-clockwise 
from the first given point ($x_1$, $y_1$) to the second given point ($x_2$, $y_2$).\\
},label=code1
   }
 \begin{lstlisting}[mathescape][firstnumber=1]
   ELLIPSE_SEGMENT (A, B, X1, Y1, X2, Y2)
 do if (A  0 or B  0)
     then return (-1, ERROR_ELLIPSE_PARAMETERS)            :DATA CHECK
            2   2     2  2              2  2     2  2
 do if (|X1 /A + Y1 /B ­ 1| >  or |X1 /A + Y1 /B ­ 1| > )
     then return (-1, ERROR_POINTS_NOT_ON_ELLIPSE)         :DATA CHECK
 do if (|X1|/A > )
     do if |X1| - A > 
        then return (-1, ERROR_INVERSE_TRIG)               :DATA CHECK
        else do if X1 < 0
              then X1  -A
              else X1  A
 do if (|X2|/A > )
     do if |X2| - A > 
        then return (-1, ERROR_INVERSE_TRIG)               :DATA CHECK
        else do if X2 < 0
                  then X2  -A
                  else X2  A
 do if (Y1 < 0)                     :ANGLE QUADRANT FORMULA (TABLE 1)
     then 1  2 ­ acos (X1/A)
     else 1  acos (X1/A)
 do if (Y2 < 0)                     :ANGLE QUADRANT FORMULA (TABLE 1)
     then 2  2 ­ acos (X2/A)
     else 2  acos (X2/A)
 do if (1 > 2)                                :MUST START WITH 1 < 2
     then 1  1 - 2
 do if ((2 ­ 1) > )                       :STORE SIGN OF TRIANGLE AREA
     then trsgn  +1.0
     else trsgn  +1.0
 area  0.5*(A*B*(2 - 1) ­ trsgn*|X1*Y2 - X2*Y1|)
 return (area, NORMAL_TERMINATION)
\end{lstlisting}
      
 An implementation of the ELLIPSE\_SEGMENT algorithm written in c--code is shown in
Appendix~\ref{appA}. The code compiles under Cygwin-1.7.7-1, and returns the following values for the
two test cases presented in Fig.~\ref{fig1} and Fig.~\ref{fig2}:

\lstset{language=FORTRAN, 
      stringstyle=\color{black},
  keywordstyle=\color{cyan},
caption={Return values for the test cases in Fig.~\ref{fig1} and Fig.~\ref{fig2}},label=code11
   }
\begin{lstlisting}[mathescape][firstnumber=1]
        cc call_es.c ellipse_segment.c -o call_es.exe
        ./call_es
       Calling ellipse_segment.c
       Fig. 1: segment area =               9.52864712, return_value = 0
       Fig. 2: segment area =             15.60409411, return_value = 0
       sum of ellipse segments = 25.13274123
       ellipse area by pi*A*B =           25.13274123
\end{lstlisting}

\section{Extending the Core Segment Algorithm to more General Cases}
\subsection{Segment Area for a (Directional) Line through a General Ellipse}

The core segment algorithm is based on an ellipse that is centered at the origin with its axes aligned to the coordinate axes.  The algorithm can be extended to more general ellipses, such as rotated and/or translated ellipse forms.  Start by considering the case for a standard ellipse with semi-major axis lengths of $A$\, and $B$ that is centered at the origin and with its axes aligned with the coordinate axes.  Suppose that the ellipse is rotated through a counter-clockwise angle $\varphi $, and that the ellipse is then translated so that its center is at the point ($h$, $k$).  The rotated+translated ellipse could then be defined by the set of parameters ($A$, $B$, $h$, $k$, $\varphi$), with the understanding that the rotation through \textit{$\varphi $} is performed before the translation through ($h$, $k$).  The approach for extending the core segment area algorithm will be to determine analogs on the standard ellipse corresponding to any points of intersection between a shape of interest and the general rotated and translated ellipse.  To identify corresponding points, features of the shape of interest are translated by (--$h$, --$k$), and then rotated by --$\varphi$.  The translated+rotated features are used to determine any points of intersection with a similar ellipse that is centered at the origin with its axes aligned to the coordinate axes.  Then, the core segment algorithm can be called with the translated+rotated intersection points.

\begin{figure}[htp]
\begin{center}
  \includegraphics[width=2in]{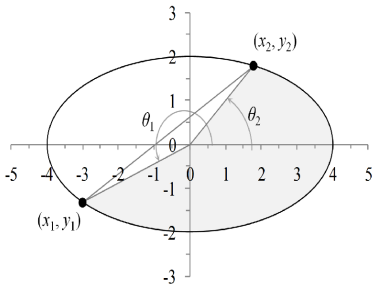}
  \caption{Translation and rotation are affine transformations that are also length-and area-preserving.  Corresponding points on the two ellipses will demarcate equal partition areas.}
\label{fig3}
  \end{center}
\end{figure}

Rotation and translation are affine transformations that are also length- and area-preserving.  In particular, the semi-axis lengths in the general rotated ellipse are preserved by both transformations, and corresponding points on the two ellipses will demarcate equal partition areas.  Fig.~\ref{fig3} illustrates this idea, showing the ellipse of Fig.~\ref{fig1} which has been rotated counter-clockwise through an angle $\varphi = 3\pi /8$, then translated by $(h, k) = (-6, +3)$.

Suppose that we desire to find the area of the rotated+translated ellipse sector defined by the line $y = -x$, where the line `direction' travels from lower-right to upper-left, as shown in Fig.~\ref{fig3}.  We describe an approach for finding a segment in a rotated+translated ellipse, based on the core ellipse segment algorithm.

An ellipse that is centered at the origin, with its axes aligned to the coordinate axes, is defined parametrically by
\[\left. \begin{array}{c}
x=A\cdot \cos(t) \\ 
y=B\cdot \sin(t) \end{array}
\right\}\ \ 0\le t\le 2\pi \] 
Suppose the ellipse is rotated through an angle $\varphi$, with counter-clockwise being positive, and that the ellipse is then to be translated to put its center is at the point ($h$, $k$).  Any point ($x$, $y$) on the standard ellipse can be rotated and translated to end up in a corresponding location on the new ellipse by using the transformation:
\begin{equation}
\left[ \begin{array}{c}
x_{TR} \\ 
y_{TR} \end{array}
\right]=\left[ \begin{array}{cc}
{\cos  \left(\varphi \right)\ } & -{\sin  \left(\varphi \right)\ } \\ 
{\sin  \left(\varphi \right)\ } & {\cos  \left(\varphi \right)\ } \end{array}
\right]\cdot \left[ \begin{array}{c}
x \\ 
y \end{array}
\right]+\left[ \begin{array}{c}
h \\ 
k \end{array}
\right]
\end{equation}
Rotation and translation of the original standard ellipse does not change the ellipse area, or the semi-axis lengths.  One important feature of the algorithms presented here is that the semi-axis lengths $A$ and $B$ are in the direction of the $x$- and \textit{y}-axes, respectively, in the \textit{un-rotated} (standard) ellipse.  In its rotated orientation, the semi-axis length $A$ will rarely be oriented horizontally (in fact, for $\varphi = \pi/4$, the semi-axis length $A$ will be oriented vertically).  Regardless of the orientation of the rotated+translated ellipse, the algorithms presented here assume that the values of $A$ and $B$ passed into the algorithm represent the semi-axis lengths along the $x$- and $y$-axes, respectively, for the corresponding un-rotated, un-translated ellipse.  The angle $\varphi$ is the amount of counter-clockwise rotation required to put the ellipse into its desired location.  Specifying a negative value for $\varphi$ will rotate the standard ellipse through a clockwise angle.  The angle $\varphi$ can be specified in anywhere in the range (--8, +8); the working angle in the code will be computed from the given angle, modulo $2\pi$, to avoid any potential errors (?) when calculating trigonometric values.  The translation ($h$, $k$) is the absolute movement along the coordinate axes of the ellipse center to move a standard ellipse into its desired location.  Negative values of $h$ move the standard ellipse to the left; negative values of $k$ move the standard ellipse down.

To find the area between the given line and the rotated+translated ellipse, the two curve equations can be solved simultaneously to find any points of intersection.  But instead of searching for the points of intersection with the rotated+translated ellipse, it is more efficient to transform the two given points that define the line back through the translation (--$h$, --$k$) then rotation through --$\varphi$.  The new line determined by the translated+rotated points will pass through the standard ellipse at points that are analogous to where the original line intersects the rotated+translated ellipse.

The transformations required to move the given points ($x_{1}$, $y_{1}$) and ($x_{2}$, $y_{2}$) into an orientation with respect to a standard ellipse that is analogous to their orientation to the given ellipse are the inverse of what it took to rotate+translate the ellipse to its desired position.  The translation is performed first, then the rotation:
\begin{equation}
\left[ \begin{array}{c}
x_{i_0} \\ 
y_{i_0} \end{array}
\right]=\left[ \begin{array}{cc}
{\cos  \left(-\varphi \right)\ } & -{\sin  \left(-\varphi \right)\ } \\ 
{\sin  \left(-\varphi \right)\ } & {\cos  \left(-\varphi \right)\ } \end{array}
\right]\cdot \left[ \begin{array}{c}
x_i-h \\ 
y_i-k \end{array}
\right]
\end{equation} 
Multiplying the vector by the matrix, and simplifying the negative-angle trig functions gives the following expressions for the translated+rotated points:
\begin{align*}
x_{i_0}=&{\cos  \left(\varphi \right)\ }\cdot \left(x_i-h\right)+{\sin  \left(\varphi \right)\cdot \left(y_i-k\right)\ }\\
y_{i_0}=&{-\sin  \left(\varphi \right)\cdot \left(x_i-h\right)\ }+{\cos  \left(\varphi \right)\ }\cdot \left(y_i-k\right)
\end{align*} 
The two new points $\left(x_{1_0},y_{1_0}\right)$ and $\left(x_{2_0},y_{2_0}\right)$ can be used to determine a line, \textit{e.g}., by the point-slope method:
\begin{equation}
y=y_{1_0}+\frac{y_{2_0}-y_{1_0}}{x_{2_0}-x_{1_0}}\left({x-x}_{1_0}\right)
\end{equation} 
The equation can also be formulated in an alternative way to accommodate cases where the translated+rotated line is vertical, or nearly so:

\begin{equation}
x=x_{1_0}+\frac{x_{2_0}-x_{1_0}}{y_{2_0}-y_{1_0}}\left({y-y}_{1_0}\right)
 \end{equation}  
Points of intersection are found by substituting the line equations into the standard ellipse equation, and solving for the remaining variable.  For each case, define the slope as:
\begin{equation}
m_{yx}=\frac{y_{2_0}-y_{1_0}}{x_{2_0}-x_{1_0}},\;\; m_{xy}=\frac{x_{2_0}-x_{1_0}}{y_{2_0}-y_{1_0}}
\end{equation}  

Then the two substitutions proceed as follows:

\begin{equation}
\begin{split}
y=&y_{1_0}+m_{yx}\cdot \left({x-x}_{1_0}\right){\rm \ \ into\ \ }\frac{x^2}{A^2}+\frac{y^2}{B^2}=1\\
\Longrightarrow& \frac{x^2}{A^2}+\frac{{\left(y_{1_0}+m_{yx}\cdot \left({x-x}_{1_0}\right)\right)}^2}{B^2}=1\\
\Longrightarrow& \left[\frac{B^2+A^2\cdot {\left(m_{yx}\right)}^2}{A^2}\right]\cdot x^2\\
+&\left[2\cdot \left(y_{1_0}\cdot m_{yx}-{\left(m_{yx}\right)}^2\cdot x_{1_0}\right)\right]\cdot x\\
+&\left[{\left(y_{1_0}\right)}^2-2\cdot m_{yx}\cdot x_{1_0}\cdot y_{1_0}+{\left(m_{yx}\cdot x_{1_0}\right)}^2-B^2\right]\\
=&0
\end{split}
\end{equation}

\begin{equation}
\begin{split}
x=&x_{1_0}+m_{xy}\cdot \left({y-y}_{1_0}\right){\rm \ \ into\ \ }\frac{x^2}{A^2}+\frac{y^2}{B^2}=1\\
\Longrightarrow& \frac{{\left(x_{1_0}+m_{xy}\cdot \left({y-y}_{1_0}\right)\right)}^2}{A^2}+\frac{y^2}{B^2}=1\\
\Longrightarrow& \left[\frac{A^2+B^2\cdot {\left(m_{xy}\right)}^2}{B^2}\right]\cdot y^2\\
+&\left[2\cdot \left(x_{1_0}\cdot m_{xy}-{\left(m_{xy}\right)}^2\cdot y_{1_0}\right)\right]\cdot y\\
+&\left[{\left(x_{1_0}\right)}^2-2\cdot m_{xy}\cdot x_{1_0}\cdot y_{1_0}+{\left(m_{xy}\cdot y_{1_0}\right)}^2-A^2\right]\\
=&0
\end{split}
\end{equation}

If the translated+rotated line is not vertical, then use the first equation to find the $x$-values for any points of intersection.  If the translated+rotated line is close to vertical, then the second equation can be used to find the $y$-values for any points of intersection.  Since points of intersection between the line and the ellipse are determined by solving a quadratic equation $ax^{2} + bx + c$, there are three cases to consider:
\begin{enumerate}
\item $\Delta=b^{2} - 4ac < 0$: Complex Conjugate Roots (no points of intersection)
\item $\Delta=b^{2} - 4ac = 0$: One Double Real Root (1 point of intersection; line tangent to ellipse)

\item $\Delta=b^{2} - 4ac > 0$: Two Real Roots (2 points of intersection; line crosses ellipse)
\end{enumerate}
 For the first two cases, the segment area will be zero.  For the third case, the two points of intersection can be sent to the core segment area algorithm.  However, to enforce a consistency in area measures returned by the core algorithm, the integration direction is specified to be from the first point to the second point.  As such, the ellipse line overlap algorithm should be sensitive to the order that the points are passed to the core segment algorithm.  We suggest giving the line a `direction' from the first given point on the line to the second.  The line `direction' can then be used to determine which is to be the first point of intersection, i.e., the first intersection point is where the line enters the ellipse based on what `direction' the line is pointing.  The segment area that will be returned from ELLIPSE\_SEGMENT by passing the line's entry location as the first intersection point is the area within the ellipse to the right of the line's path.

The approach outlined above for finding the overlap area between a line and a general ellipse is implemented in the ELLIPSE\_LINE\_OVERLAP algorithm, with pseudo-code shown in List.~\ref{code2}.  The ellipse is passed to the algorithm by specifying the counterclockwise rotation angle $\varphi$ and the translation ($h$, $k$) that takes a standard ellipse and moves it to the desired orientation, along with the semi-axes lengths, $A > 0$ and $> 0$.  The line is passed to the algorithm as two points on the line, ($x_{1}$, $y_{1}$) and ($x_{2}$, $y_{2}$).  The `direction' of the line is taken to be from ($x_{1}$, $y_{1}$) toward ($x_{2}$, $y_{2}$).  Then, the segment area returned from ELLIPSE\_SEGMENT will be the area within the ellipse to the right of the line's path.

\lstset{language=FORTRAN, 
      stringstyle=\color{black},
  keywordstyle=\color{cyan},
caption={The ELLIPSE\_LINE\_OVERLAP algorithm is shown for calculating the area of a segment in a general ellipse that is defined by a given line.  The line is considered to have a `direction' that runs from the first given point ($x_{1}$, $y_{1}$) to the second given point ($x_{2}$, $y_{2}$).  The line `direction' determines the order in which intersection points are passed to the ELLIPSE\_SEGMENT algorithm, which will return the area of the segment that runs along the ellipse from the first point to the second in a counter-clockwise direction.  Any routine that calls the algorithm ELLIPSE\_LINE\_OVERLAP must be sensitive to the order of points that are passed in.\\}
,label=code2
   }
\begin{lstlisting}[mathescape][firstnumber=1]
 (Area,Code) $\leftarrow$ ELLIPSE\_LINE\_OVERLAP (A,B,H,K,$\varphi $,X1,Y1,X2,Y2)
  do if (A $\le$ 0 or B $\le$ 0)

     then return (-1, ERROR_ELLIPSE_PARAMETERS)     :DATA_CHECK

  do if ( $|\varphi|  > 2\pi $)

     then $\varphi \leftarrow (\varphi\;  \text{modulo}\; 2\pi )$ :BRING $\varphi $ INTO $-2\pi \le \varphi < 2\pi $ (?)

  do if ($|X1| /A > 2\pi $)

     then X1 $\leftarrow$ -A

  X10 $\leftarrow \cos(\varphi )*(X1 - H) + \sin(\varphi )*(Y1 - K)$

  Y10 $\leftarrow -\sin(\varphi )*(X1 - H) + \cos(\varphi )*(Y1 - K)$

  X20 $\leftarrow \cos(\varphi)*(X2 - H) + \sin(\varphi )*(Y2 - K)$

  Y20 $\leftarrow -\sin(\varphi )*(X2 - H) + \cos(\varphi )*(Y2 - K)$

  do if ($|X20 - X10|  > \varepsilon$ )                            :LINE IS NOT VERTICAL

     then m $\leftarrow$ (Y20 - Y10)/(X20 - X10)  :STORE QUADRATIC COEFFICIENTS

          a $\leftarrow$ (B${}^{2}$ + (A*m)${}^{2}$)/$A^{2}$

          b $\leftarrow$ (2.0*(Y10*m -- $m^{2}$*X10))

          c $\leftarrow$ (Y10${}^{2}$ - 2.0*m*Y10*X10 + (m*X10)${}^{2}$ -- B${}^{2}$)

     else if (|Y20 -- Y10|  $> \varepsilon $)                :LINE IS NOT HORIZONTAL

          then m $\leftarrow$ (X20 - X10)/(Y20 - Y10)    :STORE QUADRATIC COEFFS

               a $\leftarrow$ (A${}^{2}$ + (B*m)${}^{2}$)/B${}^{2}$

               b $\leftarrow$ (2.0*(X10*m -- m${}^{2}$*Y10))

               c $\leftarrow$(X10${}^{2}$ - 2.0*m*Y10*X10 + (m*Y10)${}^{2}$ -- A${}^{2}$)

     else return (-1, ERROR_LINE_POINTS)      :LINE POINTS TOO $\text{CLOSE}$

  discrim $\leftarrow$ b${}^{2}$ - 4.0*a*c

  do if (discrim $<$ 0.0)                 :LINE DOES NOT CROSS ELLIPSE

     then return (0, NO_INTERSECT)

     else if (discrim $>$ 0.0)                :TWO INTERSECTION POINTS

     then root1 $\leftarrow$ (-b - sqrt (discrim))/(2.0*a)

          root2 $\leftarrow$ (-b + sqrt (discrim))/(2.0*a)

     else return (0, TANGENT)               :LINE TANGENT $\text{TO}$ ELLIPSE

  do if ($| X20 - X10|  > \varepsilon $)                        :ROOTS ARE X-VALUES

     then do if (X10 $<$ X20)        :ORDER PTS SAME AS LINE DIRECTION

             then x1 $\leftarrow$ root1

                  x2 $\leftarrow$ root2

             else x1 $\leftarrow$ root2

                  x2 $\leftarrow$ root1

     else do if (Y10 $<$ Y20)                      :ROOTS ARE Y-VALUES

             then y1 $\leftarrow$ root1       :ORDER PTS SAME AS LINE DIRECTION

                  y2 $\leftarrow$ root2

             else y1 $\leftarrow$ root2

                  y2 $\leftarrow$ root1

  (Area,Code) $\leftarrow$ ELLIPSE_SEGMENT (A,B,x1,y1,x2,y2)

  do if (Code $<$ NORMAL_TERMINATION)

     then return (-1.0, Code)

     else return (Area, TWO_INTERSECTION_POINTS)
\end{lstlisting}

An implementation of the ELLIPSE\_LINE\_OVERLAP algorithm in c-code is shown in Appendix~\ref{appB}.  The code compiles under Cygwin-1.7.7-1, and returns the following values for the test cases presented above in Fig.~\ref{fig3}, with both line `directions':
\lstset{language=FORTRAN, 
      stringstyle=\color{black},
  keywordstyle=\color{cyan},
caption={Return values for the test cases in Fig.~\ref{fig3}.\\}
,label=code22
   }
\begin{lstlisting}[mathescape][firstnumber=1]
 cc call_el.c ellipse_line_overlap.c ellipse_segment.c -o call_el.exe

 ./call_el

Calling ellipse_line_overlap.c

 area =                    4.07186819, return_value = 102

 reverse: area =           21.06087304, return_value = 102

 sum of ellipse segments =       25.13274123

 total ellipse area by pi*A*B =  25.13274123
\end{lstlisting}
\subsection{Ellipse-Ellipse Overlap Area}

\begin{figure}[htp]
\begin{center}
  \includegraphics[width=2in]{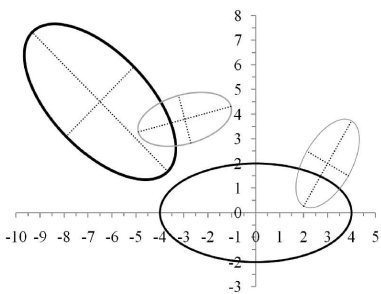}
  \caption{Intersection points on each curve are used with the ellipse segment area algorithm to determine overlap area, by calculating the area of appropriate segments, and polygons in certain cases.  For the case of two intersection points, as shown above, the overlap area can be found by adding two segments, as shown in Fig.~\ref{fig5}.}
\label{fig4}
  \end{center}
\end{figure}

The method described above for determining the area between a line and an ellipse can be extended to the task of finding the overlap area between two general ellipses.  Suppose the two ellipses are defined by their semi-axis lengths, center locations and axis rotation angles.  Let the two sets of parameters ($A_{1}$, $B_{1}$, $h_{1}$, $k_{1}$, $\varphi_{1}$) and ($A_{2}$, $B_{2}$, $h_{2}$, $k_{2}$, $\varphi_{2}$) define the two ellipses for which overlap area is sought.  The approach presented here will be to first translate both ellipses by an amount (--$h_{1}$, --$k_{1}$) that puts the center of the first ellipse at the origin.  Then, both translated ellipses are rotated about the origin by an angle --$\varphi_{1}$ that aligns the axes of the first ellipse with the coordinate axes; see Fig.~\ref{fig4}.  Intersection points are found for the two translated+rotated ellipses, using Ferrari's quartic formula.  Finally, the segment algorithm described above is employed to find all the pieces of the overlap area.

\begin{figure}[htp]
\begin{center}
  \includegraphics[width=2in]{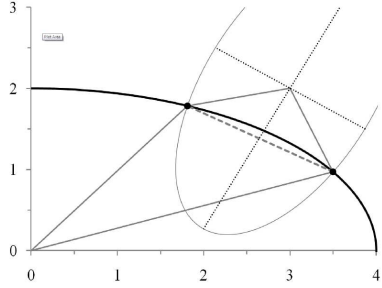}
  \caption{The area of overlap between two intersecting ellipses can be found by using the ellipse sector algorithm.  In the case of two (non-tangential) intersection points, the overlap area is equal to the sum of two ellipse sectors.  The sector in each ellipse is demarcated by the intersection points.}
\label{fig5}
  \end{center}
\end{figure}

For example, consider a case of two general ellipses with two (non-tangential) points of intersection, as shown in Fig.~\ref{fig4}.  The translation+rotation transformations that put the first ellipse at the origin and aligned with the coordinate axes do not alter the overlap area.  In the case shown in Fig.~\ref{fig4}, the overlap area consists of one segment from the first ellipse and one segment from the second ellipse.  The segment algorithm presented above can be used directly for ellipses centered at the origin and aligned with the coordinate axes.  As such, the desired segment from the first ellipse can be found immediately with the segment algorithm, based on the points of intersection.  To find the desired segment of the second ellipse, the approach presented here further translates and rotates the second ellipse so that the segment algorithm can also be used directly.  The overlap area for the case shown in Fig.~\ref{fig4} is equal to the sum of the two segment areas, as shown in Fig.~\ref{fig5}.  Other cases, e.g. with 3 and 4 points of intersection, can also be handled using the segment algorithm.

The overlap area algorithm presented here finds the area of appropriate sector(s) of each ellipse, which are demarcated by any points of intersection between the two ellipse curves.  To find intersection points, the two ellipse equations are solved simultaneously.  This step can be accomplished by using the implicit polynomial forms for each ellipse.  The first ellipse equation, in its translated+rotated position is written as an implicit polynomial using the appropriate semi-axis lengths:

\begin{equation}
\frac{x^2}{A^2_1}+\frac{y^2}{B^2_1}=1
\label{eq1}
\end{equation}

In a general form of this problem, the translation+rotation that puts the first ellipse centered at the origin and oriented with the coordinate axes will typically leave the second ellipse displaced and rotated.  The implicit polynomial form for a more general ellipse that is rotated and/or translated away from the origin is written in the conventional way as:

\begin{equation}
AA\cdot x^2+BB\cdot x\cdot y+CC\cdot y^2+DD\cdot x+EE\cdot y+FF=0
\label{eq2}
\end{equation}

Any points of intersection for the two ellipses will satisfy these two equations simultaneously.  An intermediate goal is to find the implicit polynomial coefficients in Ellipse Eq.~\ref{eq2} that describe the second ellipse after the translation+rotation that puts the first ellipse centered at the origin and oriented with the coordinate axes.  The parameters that describe the second ellipse after the translation+rotation can be determined from the original parameters for the two ellipses.  The first step is to translate the second ellipse center ($h_{2}$, $k_{2}$) through an amount (--$h_{1}$, --$k_{1}$), then rotate the center-point through --$\varphi_{1}$ to give a new center point ($h_{2TR}$, $k_{2TR}$):
\begin{align*}
h_{2TR}=&{\cos  \left({-\varphi }_1\right)\ }\cdot \left(h_2-h_1\right)-{\sin  \left({-\varphi }_1\right)\cdot \left(k_2-k_1\right)\ }\ \\ 
k_{2TR}=&{\sin  \left(-{\varphi }_1\right)\cdot \left(h_2-h_1\right)\ }+{\cos  \left({-\varphi }_1\right)\ }\cdot \left(k_2-k_1\right)\ 
 \end{align*}
The coordinates for a point ($x_{TR}$, $y_{TR}$) on the second ellipse in its new translated+rotated position can be found from the following parametric equations, based on an ellipse with semi-axis lengths $A_{2}$ and $B_{2}$ that is centered at the origin, then rotated and translated to the desired position:

\begin{equation*}
\left. \begin{array}{c}
x_{TR}=A_2\cdot \cos\left(t\right)\cdot {\cos  \left({\varphi }_2-{\varphi }_1\right)\ }-B_2\cdot \sin\left(t\right)\cdot {\sin  \left({\varphi }_2-{\varphi }_1\right)\ }+h_{2_{TR}} \\ 
y_{TR}=A_2\cdot \cos\left(t\right)\cdot {\sin  \left({\varphi }_2-{\varphi }_1\right)\ }+B_2\cdot \sin\left(t\right)\cdot {\cos  \left({\varphi }_2-{\varphi }_1\right)\ }+k_{2_{TR}} \end{array}
\right\}\ \ 0\le t\le 2\pi
\end{equation*}

To find the implicit polynomial coefficients from the parametric form, further transform the locus of points (\textit{x}${}_{TR}$, \textit{y}${}_{TR}$) so that they lie on the ellipse ($A_{2}$, $B_{2}$, 0, 0, 0), which is accomplished by first translating ($x_{TR}$, $y_{TR}$) through (--($h_{1}$ -- $h_{2}$), --($k_{1}$ -- $k_{2}$)) and then rotating the point through the angle --($\varphi_{1}$ -- $\varphi_{2}$):

 \begin{equation*}
\begin{split}
x=&{\cos  \left({\varphi }_2-{\varphi }_1\right)\ }\cdot \left(x_{TR}-\left(h_1-h_2\right)\right)-{\sin  \left({\varphi }_2-{\varphi }_1\right)\cdot \left(y_{TR}-\left(k_1-k_2\right)\right)\ }\ \\
y=&{\sin  \left({\varphi }_2-{\varphi }_1\right)\cdot \left(x_{TR}-\left(h_1-h_2\right)\right)\ }+{\cos  \left({\varphi }_2-{\varphi }_1\right)\ }\cdot \left(y_{TR}-\left(k_1-k_2\right)\right)\
\end{split}
  \end{equation*}

The locus of points ($x$, $y$) should satisfy the standard ellipse equation with the appropriate semi-axis lengths:

 \begin{equation}
\frac{x^2}{A^2_2}+\frac{y^2}{B^2_2}=1
 \end{equation}

Finally, the implicit polynomial coefficients for Ellipse Eq.~\ref{eq2} are found by substituting the expressions for the point ($x$, $y$) into the standard ellipse equation, yielding the following ellipse equation:

\begin{equation}
\begin{split}
&\frac{{\left[{\cos  \left({\varphi }_2-{\varphi }_1\right)\ }\cdot \left(x_{TR}-\left(h_1-h_2\right)\right)-{\sin  \left({\varphi }_2-{\varphi }_1\right)\cdot \left(y_{TR}-\left(k_1-k_2\right)\right)\ }\right]}^2}{A^2_2}\\ 
+&\frac{{\left[{\sin  \left({\varphi }_2-{\varphi }_1\right)\cdot \left(x_{TR}-\left(h_1-h_2\right)\right)\ }+{\cos  \left({\varphi }_2-{\varphi }_1\right)\ }\cdot \left(y_{TR}-\left(k_1-k_2\right)\right)\right]}^2}{B^2_2}\\
=&1 
\end{split}
\end{equation}

where ($xx_{TR}$, $y_{TR}$) are defined as above.  Expanding the terms, and then re-arranging the order to isolate like terms yields the following expressions for the implicit polynomial coefficients of a general ellipse with the set of parameters ($A_{2}$, $B_{2}$, $h_{2TR}$, $k_{2TR}$, $\varphi_{2}$ -- $\varphi_{1}$):\\

\begin{equation}
\begin{split}
AA=&\frac{{{\cos }^2 \left({\varphi }_2-{\varphi }_1\right)\ }}{A^2_2}+\frac{{{\sin }^2 \left({\varphi }_2-{\varphi }_1\right)\ }}{B^2_2} \\
BB=&\frac{2\cdot {\sin  \left({\varphi }_2-{\varphi }_1\right)\cdot {\cos  \left({\varphi }_2-{\varphi }_1\right)\ }\ }}{A^2_2}-\frac{2\cdot {\sin  \left({\varphi }_2-{\varphi }_1\right)\cdot {\cos  \left({\varphi }_2-{\varphi }_1\right)\ }\ }}{B^2_2} \\
CC=&\frac{{{\sin }^2 \left({\varphi }_2-{\varphi }_1\right)\ }}{A^2_2}+\frac{{{\cos }^2 \left({\varphi }_2-{\varphi }_1\right)\ }}{B^2_2} \\
DD=&\frac{-2\cdot {\cos  \left({\varphi }_2-{\varphi }_1\right)\cdot \left[h_{2_{TR}}\cdot {\cos  \left({\varphi }_2-{\varphi }_1\right)+k_{2_{TR}}\cdot {\sin  \left({\varphi }_2-{\varphi }_1\right)\ }\ }\right]\ }}{A^2_2}\\
+&\frac{2\cdot {\sin  \left({\varphi }_2-{\varphi }_1\right)\cdot \left[k_{2_{TR}}\cdot {\cos  \left({\varphi }_2-{\varphi }_1\right)\ }-h_{2_{TR}}\cdot {\sin  \left({\varphi }_2-{\varphi }_1\right)\ }\right]\ }}{B^2_2} \\
EE=&\frac{-2\cdot {\sin  \left({\varphi }_2-{\varphi }_1\right)\cdot \left[h_{2_{TR}}\cdot {\cos  \left({\varphi }_2-{\varphi }_1\right)+k_{2_{TR}}\cdot {\sin  \left({\varphi }_2-{\varphi }_1\right)\ }\ }\right]\ }}{A^2_2}\\
+&\frac{2\cdot {\cos  \left({\varphi }_2-{\varphi }_1\right)\cdot \left[h_{2_{TR}}\cdot {\sin  \left({\varphi }_2-{\varphi }_1\right)\ }-k_{2_{TR}}\cdot {\cos  \left({\varphi }_2-{\varphi }_1\right)\ }\right]\ }}{B^2_2} \\
FF=&\frac{{\left[h_{2_{TR}}\cdot {\cos  \left({\varphi }_2-{\varphi }_1\right)+k_{2_{TR}}\cdot {\sin  \left({\varphi }_2-{\varphi }_1\right)\ }\ }\right]}^2}{A^2_2}\\
+&\frac{{\left[h_{2_{TR}}\cdot {\sin  \left({\varphi }_2-{\varphi }_1\right)\ }-k_{2_{TR}}\cdot {\cos  \left({\varphi }_2-{\varphi }_1\right)\ }\right]}^2}{B^2_2}-1
\end{split}
\end{equation}

For the area overlap algorithm presented in this paper, the points of intersection between the two general ellipses are found by solving simultaneously the two implicit polynomials denoted above as Ellipse Eq.~\ref{eq1} and Ellipse Eq.~\ref{eq2}.  Solving for $x$ in the first equation:

\begin{equation}
\frac{x^2}{A^2_1}+\frac{y^2}{B^2_1}=1\ \ \ \Longrightarrow \ \ \ x=\pm \sqrt{A^2_1\cdot \left(1-\frac{y^2}{B^2_1}\right)}
\end{equation}
Substituting these expressions for $x$ into Ellipse Eq.~\ref{eq2} and then collecting terms yields a quartic polynomial in $y$.  It turns out that substituting either the positive or the negative root gives the same quartic polynomial coefficients, which are:

\begin{equation}
cy\left[4\right]\cdot y^4+cy\left[3\right]\cdot y^3+cy\left[2\right]\cdot y^2+cy\left[1\right]\cdot y+cy\left[0\right]=0
\end{equation}

where:

\begin{equation}
\begin{split}
\frac{cy\left[4\right]}{B_1}=&{A^4_1\cdot AA}^2+B^2_1\cdot \left[A^2_1\cdot \left({BB}^2-2\cdot AA\cdot CC\right)+B^2_1\cdot {CC}^2\right] \\
\frac{cy\left[3\right]}{B_1}=&2\cdot B_1\cdot \left[B^2_1\cdot CC\cdot EE+A^2_1\cdot \left(BB\cdot DD-AA\cdot EE\right)\right] \\
\frac{cy\left[2\right]}{B_1}=&A^2_1\cdot \left\{\left[B^2_1\cdot \left(2\cdot AA\cdot CC-{BB}^2\right)+{DD}^2-2\cdot AA\cdot FF\right]-2\cdot {A^2_1\cdot AA}^2\right\}\\
+&B^2_1\cdot \left(2\cdot CC\cdot FF+{EE}^2\right) \\ 
\frac{cy\left[1\right]}{B_1}=&2\cdot B_1\cdot \left[A^2_1\cdot \left(AA\cdot EE-BB\cdot DD\right)+EE\cdot FF\right] \\
\frac{cy\left[0\right]}{B_1}=&\left[A_1\cdot \left(A_1\cdot AA-DD\right)+FF\right]\cdot \left[A_1\cdot \left(A_1\cdot AA+DD\right)+FF\right] 
\end{split}
\end{equation}

In theory, the quartic polynomial will have real roots if and only if the two curves intersect.  If the ellipses do not intersect, then the quartic will have only complex roots.  Furthermore, any real roots of the quartic polynomial will represent \textit{y}-values of intersection points between the two ellipse curves.  As with the quadratic equation that arises in the ellipse-line overlap calculation, the ellipse-ellipse overlap algorithm should handle all possible cases for the types of quartic polynomial roots:
\begin{enumerate}
\item Four real roots (distinct or not); the ellipse curves intersect.

\item Two real roots (distinct or not) and one complex-conjugate pair; the ellipse curves intersect.

 \item No real roots (two complex-conjugate pairs); the ellipse curves do not intersect.
\end{enumerate}

For the method we present here, polynomial roots are found using Ferrari's quartic formula.  A numerical implementation of Ferrari's formula is given in \cite{Nonweiler}.  For complex roots are returned, and any roots whose imaginary part is returned as zero is a real root.

When the polynomial coefficients are constructed as shown above, the general case of two distinct ellipses typically results in a quartic polynomial, i.e., the coefficient $cy[4]$ is non-zero.  However, certain cases lead to polynomials of lesser degree.  Fortunately, the solver in \cite{Nonweiler} is conveniently modular, providing separate functions BIQUADROOTS, CUBICROOTS and QUADROOTS to handle all the possible polynomial cases that arise when seeking points of intersection for two ellipses.

\begin{figure}[htp]
\begin{center}
  \includegraphics[width=2in]{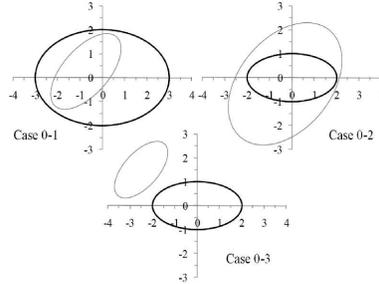}
  \caption{When the quartic polynomial has no real roots, the ellipse curves do not intersect.  It follows that either one ellipse is fully contained within the other, or the ellipse areas are completely disjoint, resulting in three distinct cases for overlap area.
}
\label{fig6}
  \end{center}
\end{figure}

If the polynomial solver returns no real roots to the polynomial, then the ellipse \textit{curves} do not intersect.  It follows that the two ellipse \textit{areas} are either disjoint, or one ellipse area is fully contained inside the other; all three possibilities are shown in Fig.~\ref{fig6}.  Each sub-case in Fig.~\ref{fig6} requires a different overlap-area calculation, i.e. either the overlap area is zero (Case 0-3), or the overlap is the area of the first ellipse (Case 0-2), or the overlap is the area of the second ellipse (Case 0-1).  When the polynomial has no real roots, geometry can be used to determine which specific sub-case of Fig.~\ref{fig6} is represented.  An efficient logic starts by determining the relative size of the two ellipses, e.g., by comparing the product of semi-axis lengths for each ellipse.  The area of an ellipse is proportional to the product of its two semi-axis lengths, so the relative size of two ellipses can be determined by comparing the product of semi-axis lengths:
\begin{equation}
(\pi \cdot A_1\cdot B_1)\; \alpha\; (\pi \cdot A_2\cdot B_2) \ \ \Longrightarrow \ \ \ (A_1\cdot B_1)\; \alpha\; (A_2\cdot B_2),\;\; \alpha \in \{'<','>'\}
\end{equation}

Suppose the first ellipse is larger than the second ellipse, then $A_{1}B_{1} >A_{2}B_{2}$.  In this case, if the second ellipse center ($h_{2TR}$, $k_{2TR}$) is inside the first ellipse, then the second ellipse is wholly contained within the first ellipse (Case 0-1); otherwise, the ellipses are disjoint (Case 0-3).  The logic relies on the fact that there are no intersection points, which is indicated whenever there are no real solutions to the quartic polynomial.  To test whether the second ellipse center ($h_{2TR}$, $k_{2TR}$) is inside the first ellipse, evaluate the first ellipse equation at the point 
$x = h_{2TR}$, and $y = k_{2TR}$; if the result is less than one, then the point ($h_{2TR}$, $k_{2TR}$) is inside the first ellipse.  The complete logic for determining overlap area when $A_{1}B_1 > A_{2}B_{2}$ is:

If the polynomial has no real roots, and $A_{1}B_{1}>A_{2}B_{2}$, and $\frac{h_{2TR }^{2} }{A_{1}^{2} } +\frac{k_{2TR }^{2} }{B_{1}^{2} } <1$, then the first ellipse wholly contains the second, otherwise the two ellipses are disjoint.

\noindent 

 Alternatively, suppose that the second ellipse is larger than the first ellipse, then $A_{1}B_{1}<A_{2}B_{2}$.  If the first ellipse center (0, 0) is inside the second ellipse, then the first ellipse is wholly contained within the second ellipse (Case 0-2); otherwise the ellipses are disjoint (Case 0-3).  Again, the logic relies on the fact that there are no intersection points,  To test whether (0, 0) is inside the second ellipse, evaluate the second ellipse equation at the origin; if the result is less than zero, then the origin is inside the second ellipse.  The complete logic for determining overlap area when $A_{1}B_{1}<A_{2}B_{2}$ is:

 If the polynomial has no real roots, and $A_{1}B_{1}<A_{2}B_{2}$, and $FF< 0$, then the second ellipse wholly contains the first, otherwise the two ellipses are disjoint.

Suppose that the two ellipses are the same size, i.e., $A_{1}B_{1}= A_{2}B_{2}$.  In this case, when no intersection points exist, the ellipses must be disjoint (Case 0-3).  It also turns out that the polynomial solver of \cite{Nonweiler} will return no real solutions if the ellipses are identical.  This special case is also handled in the overlap area algorithm presented below.  Pseudo-code for a function NOINTPTS that determines overlap area for the cases depicted in Fig.~\ref{fig6}  is shown in Fig. 14.

If the polynomial solver returns either two or four real roots to the quartic equation, then the ellipse curves intersect.  For the algorithm presented here, all of the various possibilities for the number and type of real roots are addressed by creating a list of distinct real roots.  The first step is to loop through the entire array of complex roots returned by the polynomial solver, and retrieve only real roots, i.e., only those roots whose imaginary component is zero.  The algorithm presented here then sorts the real roots, allowing for an efficient check for multiple roots.  As the sorted list of real roots is traversed, any root that is `identical' to the previous root can be skipped.

Each distinct real root of the polynomial represents a $y$-value where the two ellipses intersect.  Each $y$-value can represent either one or two potential points of intersection.  In the first case, suppose that the polynomial root is $y  = B_{1}$ (or $y = -B_{1}$), then the $y$-value produces a single intersection point, which is at (0, $B_{1}$) (or (0, -$B_{1}$)).  In the second case, if the $y$-value is in the open interval ($-B_{1}$, $B_{1}$), then there are two potential intersection points where the $y$-value is on the first ellipse:

\begin{align*}
&\left(A_1\cdot \sqrt{1-\frac{y^2}{B^2_1}},\ y\right) {\rm  and}\\
&\left({-A}_1\cdot \sqrt{1-\frac{y^2}{B^2_1}},\ y\right)
\end{align*}

Each potential intersection point ($x_{i}$, $y_{i}$) is evaluated in the second ellipse equation:

\begin{equation*}
AA\cdot x^2_i+BB\cdot x_i\cdot y_i+CC\cdot y^2_i+DD\cdot x_i+EE\cdot y_i+FF,\ \ i=1,2
\end{equation*}

\begin{figure}[htp]
\begin{center}
  \includegraphics[width=2in]{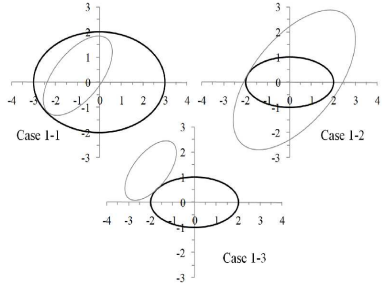}
  \caption{When only one intersection point exists, the ellipses must be tangent at the intersection point.  As with the case of zero intersection points, either one ellipse is fully contained within the other, or the ellipse areas are disjoint.  The algorithm for finding overlap area in the case of zero intersection points can also be used when there is a single intersection point.}
\label{fig7}
  \end{center}
\end{figure}

If the expression evaluates to zero, then the point ($x$, $y$) is on both ellipses, i.e., it is an intersection point.  By checking all points ($x$, $y$) for each value of $y$ that is a root of the polynomial, a list of distinct intersection points is generated.  The number of distinct intersection points must be either 0, 1, 2, 3 or 4.  The case of zero intersection points is described above, with all possible sub-cases illustrated in Fig.~\ref{fig6}.  If there is only one distinct intersection point, then the two ellipses must be tangent at that point.  The three possibilities for a single tangent point are shown in Fig.~\ref{fig7}.
 
\begin{figure}[htp]
\begin{center}
  \includegraphics[width=2in]{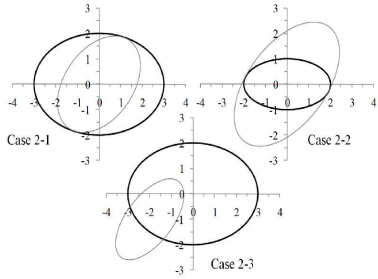}
  \caption{When two intersection points exist, either both of the points are tangents, or the ellipse curves cross at both points.  For two tangent points, one ellipse must be fully contained within the other.  For two crossing points, a partial overlap must exist}
\label{fig8}
  \end{center}
\end{figure}

For the purpose of determining overlap area, the cases of 0 or 1 intersection points can be handled in the same way.  When two intersection points exist, there are three possible sub-cases, shown in Fig.~\ref{fig8}.  It is possible that both of the intersection points are tangents (Case 2-1 and Case 2-2).  In both of these sub-cases, one ellipse must be fully contained within the other.  The only other possibility for two intersection points is a partial overlap (Case 2-3).

 Each sub-case in Fig.~\ref{fig8} requires a different overlap-area calculation.  When two intersection points exist, either both of the points are tangents, or the ellipse curves cross at both points.  Specifically, when there are two intersection points, if one point is a tangent, then both points must be tangents.  And, if one point is not a tangent, then neither point is a tangent.  So, it suffices to check one of the intersection points for tangency.  Suppose the ellipses are tangent at an intersection point; then, points that lie along the first ellipse on either side of the intersection will lie in the same region of the second ellipse (inside or outside).  That is, if two points are chosen that lie on the first ellipse, one on each side of the intersection, then both points will either be inside the second ellipse, or they will both be outside the second ellipse.  If the ellipse curves cross at the intersection point, then the two chosen points will be in different regions of the second ellipse.

A logic based on testing points that are adjacent to a tangent point can be implemented numerically to test whether an intersection point is a tangent or a cross-point.  Starting with an intersection point ($x$, $y$), calculate the parametric angle on the first ellipse, by the rules in Table~\ref{tab1}:

\begin{equation}
 \theta=
 \begin{cases}
 \arccos(x/A_1) & y\ge 0 \\ 
 2\pi -\arccos(x/A_1) & y<0
 \end{cases}
\end{equation}

A small perturbation angle is then calculated.  For the method presented here, we seek to establish an angle that corresponds to a point on the first ellipse that is a given distance, approximately $2EPS$, away from the intersection point:
\begin{equation}
{EPS}_{{\rm Radian}}={\arcsin}\left(\frac{2\cdot {\rm EPS}}{\sqrt{x^2+y^{2}}}\right)
\end{equation} 
The angle ${EPS}_{Radian}$ is then used with the parametric form of the first ellipse to determine two points adjacent to ($x$, $y$):

\begin{equation}
\begin{split}
x_1=&A_1\cdot \cos(\theta +EPS_{\rm Radian})\\
y_1=&B_1\cdot \sin(\theta +EPS_{\rm Radian})\\
x_2=&A_1\cdot \cos(\theta -EPS_{\rm Radian})\\ 
y_2=&B_1\cdot \sin(\theta -EPS_{\rm Radian})
\end{split}
\end{equation}

Each of the points is then evaluated in the second ellipse equation:
\begin{equation}
\text{test}_i=AA\cdot x^2_i+BB\cdot x_i\cdot y_i+CC\cdot y^2_i+DD\cdot x_i+EE\cdot y_i+FF,\ \ i=1,2
\end{equation}
 
If the value of $\text{test}_{i}$ is positive, then the point ($x_{i}$, $y_{i}$) is outside the second ellipse.  It follows that the product of the two test-point evaluations $\text{test}_{1}\text{test}_{2}$ will be positive if the intersection point is a tangent, since at a tangent point both test points will be on the same side of the ellipse.  The product of the test-point evaluations will be negative if the two ellipse curves cross at the intersection point, since the test points will be on opposite sides of the ellipse.  The function ISTANPT implements this logic to check whether an intersection point is a tangent or a cross-point; pseudo-code is shown in Fig. 18.

When there are two intersection points, the ISTANPT function can be used to differentiate the case 2-3 (Fig.~\ref{fig8}) from the cases 2-1 and 2-2.  Either of the two known intersection points can be checked with ISTANPT.  If the intersection point is a tangent, then both of the intersection points must be tangents, so the case is either 2-1 or 2-2, and one ellipse must be fully contained within the other.  For cases 2-1 and 2-2, the geometric logic used for 0 or 1 intersection points can also be used, i.e., the function NOINTPTS can be used to determine the overlap area for these cases.  If the two ellipse curves cross at the tested intersection point, then the case must be 2-3, representing a partial overlap between the two ellipse areas.

For case 2-3, with partial overlap between the two ellipses, the approach for finding overlap area is based on using the two points (${x}_{1}$, $y_{1}$) and ($x_{2}$, $y_{2}$) with segment the algorithm (Table 2; Fig. ~\ref{fig2}) to determine the partial overlap area contributed by each ellipse.  The total overlap area is the sum of the two segment areas.  The two intersection points divide each ellipse into two segment areas (see Fig.~\ref{fig5}).  Only one sector area from each ellipse contributes to the overlap area.  The segment algorithm returns the area between the secant line and the portion of the ellipse from the first point to the second point traversed in a counter-clockwise direction.  For the overlap area calculation, the two points must be passed to the segment algorithm in the order that will return the correct segment area.  The default order is counter-clockwise from the first point ($x_{1}$, $y_{1}$) to the second point ($x_{2}$, $y_{2}$).  A check is made to determine whether this order will return the desired segment area.  First, the parametric angles corresponding to ($x_{1}$, $y_{1}$) and ($x_{2}$, $y_{2}$) on the first ellipse are determined, by the rules in Table~\ref{tab1}:

\begin{equation}
 \theta_1=
 \begin{cases}
 \arccos(x_1/A_1) & y_1\ge 0 \\
 2\pi -\arccos(x_1/A_1) & y_1<0\\
 \end{cases}
\end{equation}

\begin{equation}
\theta_2=
\begin{cases}
\arccos(x_2/A_1) & y_2\ge 0 \\ 
2\pi -\arccos(x_2/A_1) & y_2<0 
\end{cases}
 \end{equation}

Then, a point between ($x_{1}$, $y_{1}$) and ($x_{2}$, $y_{2}$) that is on the first ellipse is found:

\begin{equation}
\begin{split}
x_{{\rm mid}}=&A_1\cdot \cos\left(\frac{\theta_1+\theta_2}{2}\right)\\ 
y_{{\rm mid}}=&B_1\cdot \sin\left(\frac{\theta_1+\theta _2}{2}\right)
\end{split}
\end{equation}
 
The point ($x_{\rm mid}$, $y_{\rm mid}$) is on the first ellipse between ($x_{1}$, $y_{1}$) and ($x_{2}$, $y_{2}$) when travelling counter- clockwise from ($x_{1}$, $y_{1}$) and ($x_{2}$, $y_{2}$).  If ($x_{\rm mid}$, $y_{\rm mid}$) is inside the second ellipse, then the desired segment of the first ellipse contains the point ($x_{\rm mid}$, $y_{\rm mid}$).  In this case, the segment algorithm should integrate in the default order, counterclockwise from ($x_{1}$,$y_{1}$) to ($x_{2}$, $y_{2}$).  Otherwise, the order of the points should be reversed before calling the segment algorithm, causing it to integrate counterclockwise from ($x_{2}$, $y_{2}$) to ($x_{1}$, $y_{1}$).  The area returned by the segment algorithm is the area contributed by the first ellipse to the partial overlap.

The desired segment from the second ellipse is found in a manner to the first ellipse segment.  A slight difference in the approach is required because the segment algorithm is implemented for ellipses that are centered at the origin and oriented with the coordinate axes; but, in the general case the intersection points ($x_{1}$, $y_{1}$) and ($x_{2}$, $y_{2}$) lie on the second ellipse that is in a displaced and rotated location.  The approach presented here translates and rotates the second ellipse to the origin so that the segment algorithm can be used.  It suffices to translate then rotate the two intersection points by amounts that put the second ellipse centered at the origin and oriented with the coordinate axes:
\begin{equation}
\begin{split}
x_{1\rm {TR}}=&(x_1-h_{2{\rm TR}})\cdot \cos(\varphi_1-\varphi_2)+(y_1-k_{2{\rm TR}})\cdot \sin(\varphi_2-\varphi_1)\\
y_{1{\rm TR}}=&(x_1-h_{2{\rm TR}})\cdot \sin(\varphi_1-\varphi_2)+(y_1-k_{2{\rm TR}})\cdot \cos(\varphi_1-\varphi_2)\\
x_{2{\rm TR}}=&(x_2-h_{2{\rm TR}})\cdot \cos(\varphi_1-\varphi_2)+(y_2-k_{2{\rm TR}})\cdot \sin(\varphi_2-\varphi_1)\\
y_{2{\rm TR}}=&(x_2-h_{2{\rm TR}})\cdot \sin(\varphi_1-\varphi_2)+(y_2-k_{2{\rm TR}})\cdot \cos(\varphi_1-\varphi_2)
\end{split}
\end{equation}
The new points ($x_{1{\rm TR}}$, $y_{1{\rm TR}}$) and ($x_{2{\rm TR}}$, $y_{2{\rm TR}}$) lie on the second ellipse after a translation+rotation that puts the second ellipse at the origin, oriented with the coordinate axes.  The new points can be used as inputs to the segment algorithm to determine the overlap area contributed by the second ellipse.  As with the first ellipse, the order of the points must be determined so that the segment algorithm returns the appropriate area.  The default order is counter-clockwise from the first point ($x_{1{\rm TR}}$, $y_{1{\rm TR}}$) to the second point ($x_{2{\rm TR}}$, $y_{2{\rm TR}}$).  A check is made to determine whether this order will return the desired segment area.  First, the parametric angles corresponding to points ($x_{1{\rm TR}}$, $y_{1{\rm TR}}$) and ($x_{2{\rm TR}}$, $y_{2{\rm TR}}$) on the second ellipse are determined, by the rules in Table~\ref{tab1}:

\begin{equation}
 \theta_1=
 \begin{cases}
 \arccos(x_{1{\rm TR}}/A_2) &y_{1{\rm TR}}\ge 0\\
 2\pi - \arccos(x_{\rm 1{\rm TR}}/A_2) & y_{1{\rm TR}}<0
 \end{cases}
\end{equation}
\begin{equation}
 \theta_2=
 \begin{cases}
 \arccos(x_{2{\rm TR}}/A_2) &y_{2{\rm TR}}\ge 0\\
 2\pi - \arccos(x_{\rm 2{\rm TR}}/A_2) & y_{2{\rm TR}}<0
 \end{cases}
\end{equation}

Then, a point on the second ellipse between ($x_{1{\rm TR}}$, $y_{1{\rm TR}}$) and ($x_{2{\rm TR}}$, $y_{2{\rm TR}}$) is found:

\begin{align*}
x_{{\rm mid}}=&A_2\cdot \cos\left(\frac{\theta_1 +\theta_2}{2}\right)\\
y_{{\rm mid}}=&B_2\cdot \sin\left(\frac{\theta_1 +\theta_2}{2}\right)
 \end{align*}

The point ($x_{\rm mid}$, $y_{\rm mid}$) is on the second ellipse between ($x_{1{\rm TR}}$, $y_{1{\rm TR}}$) and ($x_{2{\rm TR}}$, $y_{2{\rm TR}}$) when travelling counter- clockwise from ($x_{1{\rm TR}}$, $y_{1{\rm TR}}$) and ($x_{2{\rm TR}}$, $y_{2{\rm TR}}$).  The new point (\textit{x}${}_{mid}$, \textit{y}${}_{mid}$) lies on the centered second ellipse.  To determine the desired segment of the second ellipse, the new point ($x_{\rm mid}$, $y_{\rm mid}$) must be rotated then translated back to a corresponding position on the once-translated+rotated second ellipse:

\begin{align*}
x_{{\rm mid}{\rm RT}}=&x_{{\rm mid}}\cdot \cos(\varphi_2-\varphi_1)+y_{{\rm mid}}\cdot \sin(\varphi_1-\varphi_2)+h_{2{\rm TR}}\\
y_{{\rm mid}{\rm RT}}=&x_{{\rm mid}}\cdot \sin(\varphi_2-\varphi_1)+y_{{\rm mid}}\cdot \cos(\varphi_1-\varphi_2)+k_{2{\rm TR}}
\end{align*}

If ($x_{{\rm mid}RT}$, $y_{{\rm mid}{\rm RT}}$) is inside the first ellipse, then the desired segment of the second ellipse contains the point ($x_{\rm mid}$, $y_{\rm mid}$).  In this case, the segment algorithm should integrate in the default order, counterclockwise from ($x_{1{\rm TR}}$, $y_{1{\rm TR}}$) to ($x_{2{\rm TR}}$, $y_{2{\rm TR}}$).  Otherwise, the order of the points should be reversed before calling the segment algorithm, causing it to integrate counterclockwise from ($x_{2{\rm TR}}$, $y_{2{\rm TR}}$) to ($x_{1{\rm TR}}$, $y_{1{\rm TR}}$).  The area returned by the segment algorithm is the area contributed by the second ellipse to the partial overlap.  The sum of the segment areas from the two ellipses is then equal to the ellipse overlap area.  The TWOINTPTS function calculates the overlap area for partial overlap with two intersection points (Case 2-3); pseudo-code is shown in Fig. 15.

\begin{figure}
\begin{center}
  \includegraphics[width=2in]{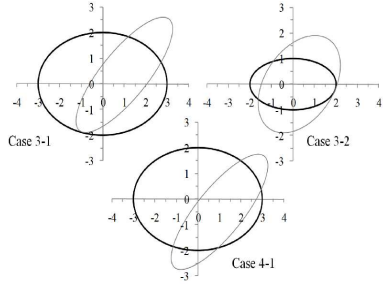}
  \caption{When three intersection points exist, one must be a tangent, and the ellipse curves must cross at the other two points, always resulting in a partial overlap.  When four intersection points exist, the ellipse curves must cross at all four points, again resulting in a partial overlap}
\label{fig9}
  \end{center}
\end{figure}

There are two possible sub-cases for three intersection points, shown in Fig.~\ref{fig9}.  One of the three points must be a tangent point, and the ellipses must cross at the other two points.  The cases are distinct only in the sense that the tangent point occurs with ellipse 2 on the interior side of ellipse 1 (Case 3-1), or with ellipse 2 on the exterior side of ellipse 1 (Case 3-2).  The overlap area calculation is performed in the same manner for both cases, by calling the TWOINTPTS function with the two cross-point intersections.  The ISTANPT function can be used to determine which point is a tangent; the remaining two intersection points are then passed to TWOINTPTS.  This logic is implemented in the THREEINTPTS function, with pseudo-code in Fig. 16.

There is only one possible case for four intersection points, shown in Fig.~\ref{fig9}.  The two ellipse curves must cross at all four of the intersection points, resulting in a partial overlap.  The overlap area consists of two segments from each ellipse, and a central convex quadrilateral.  For the approach presented here, the four intersection points are sorted ascending in a counter-clockwise order around the first ellipse.  The ordered set of intersection points is ($x_{1}$, $y_{1}$), ($x_{2}$, $y_{2}$), ($x_{3}$, $y_{3}$) and ($x_{4}$, $y_{4}$).  The ordering allows a direct calculation of the quadrilateral area.  The standard formula uses the cross-product of the two diagonals:
\begin{equation}
\begin{split}
{\rm area}=&\frac{1}{2}\left|\left(x_3-x_1,y_3-y_1\right)\times \left(x_4-x_2,y_4-y_2\right)\right|\\
=&\frac{1}{2}\left|\left(x_3-x_1\right)\cdot \left(y_4-y_2\right)-\left(x_4-x_2\right)\cdot \left(x_3-x_1\right)\right|
\end{split}
\end{equation}
  
The point ordering also simplifies the search for the appropriate segments of each ellipse that contribute to the overlap area.

\begin{figure}[htp]
\begin{center}
  \includegraphics[width=2in]{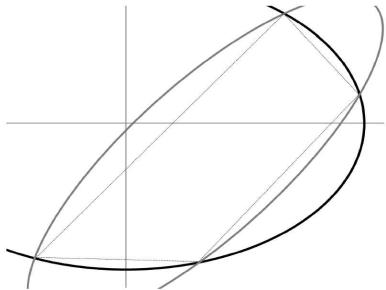}
  \caption{Overlap Area with four intersection points (Case 4-1).  The overlap area consists of two segments from each ellipse, and a central convex quadrilateral.}
\label{fig10}
  \end{center}
\end{figure}

Suppose that the first two sorted points ($x_{1}$, $y_{1}$)  and ($x_{2}$, $y_{2}$) demarcate a segment of the first ellipse that contributes to the overlap area, as shown in  Fig.~\ref{fig9} and Fig.~\ref{fig10}. It follows that the contributing segments from the first ellipse are between ($x_{1}$, $y_{1}$) and ($x_{2}$, $y_{2}$), and also between ($x_{3}$, $y_{3}$) and ($x_{4}$, $y_{4}$).  In this case, the contributing segments from the second ellipse are between ($x_{2}$, $y_{2}$) and ($x_{3}$, $y_{3}$), and between ($x_{4}$, $y_{4}$) and ($x_{1}$, $y_{1}$).  To determine which segments contribute to the overlap area, it suffices to test whether a point midway between ($x_{1}$, $y_{1}$) and ($x_{2}$, $y_{2}$) is inside or outside the second ellipse. The segment algorithm is used for each of the four areas, and added to the quadrilateral to obtain the total overlap area.

An implementation of the ELLIPSE\_ELLIPSE\_OVERLAP algorithm in c-code is shown in Appendix\ref{appC}.  The code compiles under Cygwin-1.7.7-1, and returns the following values for the test cases presented above in Fig.~\ref{fig6}, Fig.~\ref{fig7} , Fig.~\ref{fig8} and Fig.~\ref{fig9}:

\lstset{language=FORTRAN, 
      stringstyle=\color{black},
  keywordstyle=\color{cyan},
caption={Return values for the test cases presented above in Fig.~\ref{fig6}, Fig.~\ref{fig7} , Fig.~\ref{fig8} and Fig.~\ref{fig9}.\\}
,label=code3
   }
\begin{lstlisting}[mathescape][firstnumber=1]
 cc call\_ee.c ellipse\_ellipse\_overlap.c -o call\_ee.exe

 ./call\_ee

 Calling ellipse\_ellipse\_overlap.c

 

 Case 0-1: area =      6.28318531, return_value = 111

           ellipse 2 area by pi*a2*b2 =      6.28318531

 Case 0-2: area =      6.28318531, return_value = 110

           ellipse 1 area by pi*a1*b1 =      6.28318531

 Case 0-3: area =      0.00000000, return_value = 103

           Ellipses are disjoint, ovelap area = 0.0

 

 Case 1-1: area =      6.28318531, return_value = 111

           ellipse 2 area by pi*a2*b2 =      6.28318531

 Case 1-2: area =      6.28318531, return_value = 110

           ellipse 1 area by pi*a1b1 =      6.28318531

 Case 1-3: area =     -0.00000000, return_value = 107

           Ellipses are disjoint, ovelap area = 0.0

 

 Case 2-1: area =     10.60055478, return_value = 109

           ellipse 2 area by pi*a2*b2 =     10.60287521

 Case 2-2: area =      6.28318531, return_value = 110

           ellipse 1 area by pi*a1b1 =      6.28318531

 Case 2-3: area =      3.82254574, return_value = 107

 

 Case 3-1: area =      7.55370392, return_value = 107

 Case 3-2: area =      5.67996234, return_value = 107

 

 Case 4-1: area =     16.93791852, return_value = 109
\end{lstlisting}

\lstset{language=FORTRAN, 
      stringstyle=\color{black},
  keywordstyle=\color{cyan},
caption={The ELLIPSE\_ELLIPSE\_OVERLAP algorithm is shown for calculating the overlap area between two general ellipses.  The algorithm calls several supporting functions, including the polynomial solvers BIQUADROOTS, CUBICROOTS and QUADROOTS, from CACM Algorithm 326 \cite{Chraibi2010a} .  The remaining functions are outlined in figures below.\\}
,label=code4
   }
\begin{lstlisting}[mathescape][firstnumber=1]

(Area,Code) $\leftarrow$ ELLIPSE_ELLIPSE_OVERLAP (A1,B1,H1,K1,$\varphi $1,A2,B2,H2,K2,$\varphi $2)

 do if (A1 = 0 or B1 = 0) OR (A2 = 0 or B2 = 0)

    then return (-1, ERROR_ELLIPSE_PARAMETERS)           :DATA CHECK

 do if ($|\varphi 1|  > 2\pi $)

    then $\varphi 1 \leftarrow (\varphi 1 \text{modulo}\, 2\pi $)

 do if ($|\varphi 2|  > 2\pi $)

    then $\varphi 2 \leftarrow (\varphi 2 \text{modulo}\, 2\pi $)

 H2_TR $\leftarrow$ (H2 - H1)*cos ($\varphi 1$) + (K2 - K1)*sin ($\varphi 1$)     :TRANS+ROT ELL2

 K2_TR $\leftarrow$ (H1 - H2)*sin ($\varphi 1$) + (K2 - K1)*cos ($\varphi 1$)

 $\varphi 2R$ $\leftarrow$ $\varphi 2$ -- $\varphi 1$

 do if ($|\varphi 2R|> 2\pi $)

    then $\varphi 2R \leftarrow (\varphi 2R modulo 2\pi $)

 AA $\leftarrow$ cos${}^{2}$($\varphi $2R)/A2${}^{2}$ + sin${}^{2}$($\varphi $2R)/B2${}^{2}$     :BUILD\, $\text{IMPLICIT}$\, COEFFS ELL2TR

 BB $\leftarrow$ 2*cos ($\varphi $2R)*sin ($\varphi $2R)/A2${}^{2}$ -- 2*cos ($\varphi $2R)*sin ($\varphi $2R)/B2${}^{2}$

 CC $\leftarrow$ sin${}^{2}$($\varphi $2R)/A2${}^{2}$ + cos${}^{2}$($\varphi $2R)/B2${}^{2}$

 DD $\leftarrow$ -2*cos ($\varphi $2R)*(cos ($\varphi $2R)*H2_TR + sin ($\varphi $2R)*K2_TR)/A2${}^{2}$

      - 2*sin ($\varphi $2R)*(sin ($\varphi $2R)*H2_TR - cos ($\varphi $2R)*K2_TR)/B2${}^{2}$

EE $\leftarrow$ -2*sin ($\varphi $2R)* (cos ($\varphi $2R)*H2_TR + sin ($\varphi $2R)*K2_TR)/A2${}^{2}$

      + 2*cos ($\varphi $2R)* (sin ($\varphi $2R)*H2_TR - cos ($\varphi $2R)*K2_TR)/B2${}^{2}$

FF $\leftarrow$ (-cos ($\varphi $2R)*H2_TR - sin ($\varphi $2R)*K2_TR)${}^{2}$/A2${}^{2}$ 

      + (sin ($\varphi $2R)*H2_TR - cos ($\varphi $2R)*K2_TR)${}^{2}$/B2${}^{2}$ - 1

    :BUILD QUARTIC POLYNOMIAL COEFFICIENTS FROM THE TWO ELLIPSE EQNS

 cy[4] $\leftarrow$ A1${}^{4}$*AA${}^{2}$ + B1${}^{2}$*(A1${}^{2}$*(BB${}^{2}$ - 2*AA*CC)+ B1${}^{2}$*CC${}^{2}$)

 cy[3] $\leftarrow$ 2*B1*(B1${}^{2}$*CC*EE + A1${}^{2}$*(BB*DD - AA*EE))

 cy[2] $\leftarrow$ A1${}^{2}$*((B1${}^{2}$*(2*AA*CC -- BB${}^{2}$) + DD${}^{2}$ - 2*AA*FF)

         -2*A1${}^{2}$*AA${}^{2}$ + B1${}^{2}$*(2*CC*FF + EE${}^{2}$)

 cy[1] $\leftarrow$ 2*B1*(A1${}^{2}$*(AA*EE -- BB*DD) + EE*FF)

 cy[0] $\leftarrow$ (A1*(A1*AA --DD) + FF)*(A1*(A1*AA + DD) + FF)

 py[0] $\leftarrow$ 1

 do if ($|cy[4]|>$ 0)                               :SOLVE QUARTIC EQ

    then for i $\leftarrow$ 0 to 3 by 1

            py[4-i] $\leftarrow$ cy[i]/cy[4]

         r[][]$\leftarrow$ BIQUADROOTS (py[])

         nroots $\leftarrow$ 4

    else if ($|cy[3]|  > 0$)                            :SOLVE CUBIC EQ

    then for i $\leftarrow$ 0 to 2 by 1

            py[3-i] $\leftarrow$ cy[i]/cy[3]

         r[][] $\leftarrow$ CUBICROOTS (py[])

         nroots $\leftarrow$ 3

    else if ($|cy[2]|> 0$)                        :SOLVE QUADRATIC EQ

    then for i $\leftarrow$ 0 to 1 by 1

            py[2-i] $\leftarrow$ cy[i]/cy[2]

         r[][] $\leftarrow$ QUADROOTS (py[])

         nroots $\leftarrow$ 2

    else if ($|cy[1]|> 0$)                           :SOLVE LINEAR EQ

    then r[1][1] $\leftarrow$ (-cy[0]/cy[1])

         r[2][1] $\leftarrow$ 0

         nroots $\leftarrow$ 1

    else                                   :COMPLETELY DEGENERATE EQ

         nroots $\leftarrow$ 0

 nychk $\leftarrow$ 0                                      :IDENTIFY $\text{REAL}$ ROOTS

 for i $\leftarrow$ 1 to nroots by 1

    do if ($|r[2][i]|< \text{EPS}$)

       then nychk $\leftarrow$ nychk + 1

            ychk[nychk] $\leftarrow$ r[1][i]*B1

 for j $\leftarrow$ 2 to nychk by 1                            :SORT $\text{REAL}$ ROOTS

    tmp0 $\leftarrow$ ychk[j]

    for k $\leftarrow$ (j -- 1) to 1 by -1

       do if (ychk[k] = tmp0)

          then break

          else ychk[k+1] $\leftarrow$ ychk[k]

    ychk[k+1] $\leftarrow$ tmp0

 nintpts $\leftarrow$ 0                               :FIND INTERSECTION POINTS

 for i $\leftarrow$ 1 to nychk by 1

    do if (($i > 1$) and ($|ychk[i] - ychk[i-1]|<$ EPS/2))

       then continue

    do if ($|ychk[i]|>$ -B1)

       then x1 $\leftarrow$ 0

       else x1$\leftarrow$? A1*sqrt (1.0 - ychk[i]${}^{2}$/B1${}^{2}$)

    x2 $\leftarrow$ -x1

    do if ($|ellipse2tr (x1,ychk[i],AA,BB,CC,DD,EE,FF)|<$ EPS/2)

       then nintpts $\leftarrow$ nintpts + 1

            do if (nintpts $>$ 4)

               then return (-1, ERROR_INTERSECTION_PTS)

            xint[nintpts] $\leftarrow$ x1

            yint[nintpts] $\leftarrow$ ychk[i]

    do if (($|ellipse2tr (x2, ychk[i],AA,BB,CC,DD,EE,FF)|<$ EPS/2)

       and ($|x2 - x1|  >$ EPS/2))

       then nintpts $\leftarrow$ nintpts + 1

            do if (nintpts $>$ 4)

               then return (-1, ERROR_INTERSECTION_PTS)

            xint[nintpts] $\leftarrow$ x1

            yint[nintpts] $\leftarrow$ ychk[i]

 switch (nintpts)        :HANDLE ALL CASES FOR \# OF INTERSECTION PTS

    case 0:

    case 1:

       (OverlapArea,Code) $\leftarrow$ NOINTPTS (A1,B1,A2,B2,H1,K1,H2,K2,AA,

           BB,CC,DD,EE,FF)

       return (OverlapArea,Code)

    case 2:

       Code $\leftarrow$ istanpt (xint[1],yint[1],A1,B1,AA,BB,CC,DD,EE,FF)

       do if (Code == TANGENT_POINT)

          then (OverlapArea,Code) $\leftarrow$ NOINTPTS (A1,B1,A2,B2,H1,K1,

                   H2,K2,AA,BB,CC,DD,EE,FF)

          else (OverlapArea,Code) $\leftarrow$ TWOINTPTS (xint[],yint[],A1,

                PHI_1,A2,B2,H2_TR,K2_TR,PHI_2,AA,BB,CC,DD,EE,FF)

       return (OverlapArea,Code)

    case 3:

       (OverlapArea,Code) $\leftarrow$ THREEINTPTS (xint,yint,A1,B1,PHI_1,

           A2,B2,H2_TR,K2_TR,PHI_2,AA,BB,CC,DD,EE,FF)

       return (OverlapArea, Code)

    case 4:

       (OverlapArea,Code) $\leftarrow$ FOURINTPTS (xint,yint,A1,B1,PHI_1, 

           A2, B2,H2_TR,K2_TR,PHI_2,AA,BB,CC,DD,EE,FF)

       return (OverlapArea,Code)
\end{lstlisting}

\lstset{language=FORTRAN, 
      stringstyle=\color{black},
  keywordstyle=\color{cyan},
caption={The NOINTPTS subroutine.  If there are either 0 or 1 intersection points, this function determines whether one ellipse is contained within the other (Cases 0-1, 0-2, 1-1 and 1-2), or if the ellipses are disjoint (Cases 0-3 and 1-3).  The function returns the appropriate overlap area, and a code describing which case was encountered.\\}
,label=code5
   }
\begin{lstlisting}[mathescape][firstnumber=1]

 (OverlapArea,Code) $\leftarrow$ NOINTPTS (A1,B1,A2,B2,H1,K1,H2_TR,K2_TR,AA,

                                BB,CC,DD,EE,FF)

  relsize $\leftarrow$ A1*B1 - A2*B2

  do if (relsize $>$ 0)

     then do if (((H2_TR*H2_TR)/(A1*A1)+(K2_TR*K2_TR)/(B1*B1)) $<$ 1.0)

             then return ($\pi $*A2*B2,ELLIPSE2_INSIDE_ELLIPSE1)

             else return (0, DISJOINT_ELLIPSES)

     else do if (relsize $<$ 0)

             then do if (FF $<$ 0)

                     then return ($\pi $*A1*B1,ELLIPSE1_INSIDE_ELLIPSE2)

                     else return (0, DISJOINT_ELLIPSES)

     else do if ((H1 = H2_TR) AND (K1 = K2_TR))

             then return ($\pi $*A1*B1, ELLIPSES_ARE_IDENTICAL)

             else return (-1, ERROR_CALCULATIONS
\end{lstlisting}

\lstset{language=FORTRAN, 
      stringstyle=\color{black},
  keywordstyle=\color{cyan},
caption={The TWOINTPTS subroutine.  If there are 2 intersection points where the ellipse curves cross (Case 2-3), this function uses the ellipse sector algorithm to determine the contribution of each ellipse to the total overlap area.  The function returns the appropriate overlap area, and a code indicating two intersection points.\\}
,label=code6
   }
\begin{lstlisting}[mathescape][firstnumber=1]
 (OverlapArea,Code) $\leftarrow$ TWOINTPTS (xint[],yint[],A1,B1,$\varphi $1,A2,B2,H2_TR,

                                 K2_TR,$\varphi $2,AA,BB,CC,DD,EE,FF)

do if ($|x[1]|>$ A1)                      :AVOID INVERSE TRIG ERRORS

   then do if (x[1] $<$ 0)

           then x[1] $\leftarrow$ -A1

           else x[1] $\leftarrow$ A1

do if (y[1] $<$ 0)            :FIND PARAMETRIC ANGLE FOR (x[1], y[1])

   then $\theta $1 $\leftarrow$ 2$\pi $ -- arccos (x[1]/A1)

   else $\theta $1 $\leftarrow$ arccos (x[1]/A1)

do if ($|x[2]|>$ A1)                      :AVOID INVERSE TRIG ERRORS

   then do if (x[2] $<$ 0)

           then x[2] $\leftarrow$ -A1

           else x[2] $\leftarrow$ A1

do if (y[2] $<$ 0)            :FIND PARAMETRIC ANGLE FOR (x[2], y[2])

   then $\theta $2 $\leftarrow$ 2$\pi $ -- arccos (x[2]/A1)

   else $\theta $2 $\leftarrow$ arccos (x[2]/A1)

do if ($\theta $1 $>$ $\theta $2)                               :$\text{GO}$ CCW FROM $\theta $1 $\text{TO}$\, $\theta $2

   then tmp $\leftarrow$ $\theta $1, $\theta $1 $\leftarrow$ $\theta $2, $\theta $2 $\leftarrow$ tmp

xmid $\leftarrow$ A1*cos (($\theta $1 + $\theta $2)/2)

ymid $\leftarrow$ B1*sin (($\theta $1 + $\theta $2)/2)

do if (AA*xmid${}^{2}$+BB*xmid*ymid+CC*ymid${}^{2}$+DD*xmid+EE*ymid+FF $>$ 0)

   then tmp $\leftarrow$ $\theta $1, $\theta $1$\leftarrow$ $\theta $2, $\theta $2 $\leftarrow$ tmp

do if ($\theta $1 $>$ $\theta $2)                    :SEGMENT ALGORITHM FOR ELLIPSE 1

   then $\theta $1 ? $\theta $1 - 2$\pi $

do if (($\theta $2 - $\theta $1) $>$ $\pi $)

   then trsign $\leftarrow$ 1

   else trsign $\leftarrow$ -1

area1 $\leftarrow$ (A1*B1*($\theta $2 - $\theta $1) + trsign*\textbar x[1]*y[2] - x[2]*y[1])\textbar /2

x1\_tr $\leftarrow$ (x[1] - H2\_TR)*cos($\varphi $1 -- $\varphi $2) + (y[1] - K2\_TR)*sin($\varphi $2 -- $\varphi $1)

y1\_tr $\leftarrow$ (x[1] - H2\_TR)*sin($\varphi $1 -- $\varphi $2) + (y[1] - K2\_TR)*cos($\varphi $1 -- $\varphi $2)

x2\_tr $\leftarrow$ (x[2] - H2\_TR)*cos($\varphi $1 -- $\varphi $2) + (y[2] - K2\_TR)*sin($\varphi $2 -- $\varphi $1)

y2\_tr ? (x[2] - H2\_TR)*sin($\varphi $1 -- $\varphi $2) + (y[2] - K2\_TR)*cos($\varphi $1 -- $\varphi $2)

do if ($|x1\_tr|>$ A2)                     :AVOID INVERSE TRIG ERRORS

   then do if (x1_tr $<$ 0)

           then x1_tr $\leftarrow$ -A2

           else x1_tr $\leftarrow$ A2

do if (y1_tr $<$ 0)          :FIND PARAMETRIC ANGLE FOR (x1_tr, y1_tr)

   then $\theta $1 $\leftarrow$ 2$\pi $ -- arccos (x1_tr/A2)

   else $\theta $1 $\leftarrow$ arccos (x1_tr/A2)

do if ($|x2_tr|>$ A2)                     :AVOID INVERSE TRIG ERRORS

   then do if (x2_tr $<$ 0)

           then x2_tr $\leftarrow$ -A2

           else x2_tr $\leftarrow$ A2

do if (y2_tr $<$ 0)         :FIND PARAMETRIC ANGLE FOR (x2_tr, y2_tr)

   then $\theta $2 $\leftarrow$ 2$\pi $ -- arccos (x2_tr/A2)

   else $\theta $2 $\leftarrow$ arccos (x2_tr/A2)

do if ($\theta $1 $>$ $\theta $2)                               :$\text{GO}$ CCW FROM $\theta $1 $\text{TO}$\, $\theta $2

   then tmp $\leftarrow$ $\theta $1, $\theta $1 $\leftarrow$ $\theta $2, $\theta $2 $\leftarrow$ tmp

xmid $\leftarrow$ A2*cos (($\theta $1 + $\theta $2)/2)

ymid $\leftarrow$ B2*sin (($\theta $1 + $\theta $2)/2)

xmid_rt = xmid*cos($\varphi $2 -- $\varphi $1) + ymid*sin($\varphi $1 -- $\varphi $2) + H2_TR

ymid_rt = xmid*sin($\varphi $2 -- $\varphi $1) + ymid*cos($\varphi $2 -- $\varphi $1) + K2_TR

do if (xmid_rt${}^{2}$/A1${}^{2}$ + ymid_rt${}^{2}$/B1${}^{2}$ $>$ 1)

   then tmp $\leftarrow$ $\theta $1, $\theta $1 $\leftarrow$ $\theta $2, $\theta $2 $\leftarrow$ tmp

do if ($\theta $1 $>$ $\theta $2)                    :SEGMENT ALGORITHM FOR ELLIPSE 2

   then $\theta $1 $\leftarrow$ $\theta $1 - 2$\pi $

do if (($\theta $2 - $\theta $1) $>$ $\pi $)

    then trsign $\leftarrow$ 1

    else trsign $\leftarrow$ -1

area2 $\leftarrow$ (A2*B2*($\theta $2 - $\theta $1) 

+ trsign*$|x1_tr*y2_tr - x2\_tr*y1_tr)|$ /2

 return (area1 + area2, TWO_INTERSECTION_POINTS)
\end{lstlisting}

\lstset{language=FORTRAN, 
      stringstyle=\color{black},
  keywordstyle=\color{cyan},
caption={ The THREEINTPTS subroutine.  When there are three intersection points, one of the points must be a tangent point, and the ellipses must cross at the other two points.  For the purpose of determining overlap area, the TWOINTPTS function can be used with the two cross-point intersections.  The ISTANPT function can be used to determine which point is a tangent; the remaining two intersection points are then passed to TWOINTPTS.  The function returns the appropriate overlap area, and a code indicating three intersection points.\\},label=code7
   }
\begin{lstlisting}[mathescape][firstnumber=1]
 OverlapArea,Code) $\leftarrow$ THREEINTPTS (xint[],yint[],A1,B1,$\varphi $1,A2,B2,H2_TR,

                                 K2_TR,$\varphi $2,AA,BB,CC,DD,EE,FF)
tanpts $\leftarrow$ 0

for i $\leftarrow$ 1 to nychk by 1

   code $\leftarrow$ ISTANPT ISTANPT (x[i],y[i],A1,B1,AA,BB,CC,DD,EE,FF)

   do if (code = TANGENT_POINT)

      then tanpts $\leftarrow$ tanpts + 1

           tanindex $\leftarrow$ i

do if NOT (tanpts = 1)

   then return (-1, ERROR_INTERSECTION_POINTS)

switch (tanindex)                   :STORE THE INTERSECTION POINTS

   case 1:                       :TANGENT POINT IS $\text{IN}$ (x[1], y[1])

      xint[1] $\leftarrow$ xint[3]

      yint[1] $\leftarrow$ yint[3]

   case 2:                       :TANGENT POINT IS $\text{IN}$ (x[2], y[2])

      xint[2] $\leftarrow$ xint[3]

      yint[2] $\leftarrow$ yint[3]

(OverlapArea,code) $\leftarrow$ TWOINTPTS (xint[],yint[],A1,B1,$\varphi $1,A2,B2,H2_TR,

                                 K2_TR,$\varphi $2,AA,BB,CC,DD,EE,FF)

return (OverlapArea,THREE_INTERSECTION_POINTS)
\end{lstlisting}
\lstset{language=FORTRAN, 
      stringstyle=\color{black},
  keywordstyle=\color{cyan},
caption={The FOURINTPTS subroutine.  When there are four intersection points, the ellipse curves must cross at all four points.  A partial overlap area exists, consisting of two segments from each ellipse and a central quadrilateral.  The function returns the appropriate overlap area, and a code indicating four intersection points.\\},label=code8
   }
\begin{lstlisting}[mathescape][firstnumber=1]
verlapArea,Code) $\leftarrow$ FOURINTPTS (xint[],yint[],A1,B1,$\varphi $1,A2,B2,H2_TR,

                               K2_TR,$\varphi $2,AA,BB,CC,DD,EE,FF)

 for i $\leftarrow$ 1 to 4 by 1                      :AVOID INVERSE TRIG ERRORS

    do if ($|xint[i]|>$ A1)

       then do if (xint[i] $<$ 0)

               then xint[i] $\leftarrow$ -A1

               else xint[i] $\leftarrow$ A1

    do if (yint[i] $<$ 0)                      :FIND PARAMETRIC ANGLES

       then $\theta $[i] $\leftarrow$ 2$\pi $ -- arccos (xint[i]/A1)

       else $\theta $[i] $\leftarrow$ arccos (xint[i]/A1)

 for j $\leftarrow$ 2 to 4 by 1                        :PUT POINTS $\text{IN}$ CCW ORDER

    tmp0 $\leftarrow$ $\theta $[j]

    tmp1 $\leftarrow$ xint[j]

    tmp2 $\leftarrow$ yint[j]

    for k $\leftarrow$ (j-1) to 1 by -1                :INSERTION SORT BY ANGLE

       do if ($\theta $[k] $<$= tmp0)

          then break

          else $\theta $[k+1] $\leftarrow$ $\theta $[k]

               xint[k+1] $\leftarrow$ xint[k]

               yint[k+1] $\leftarrow$ yint[k]

 area1 $\leftarrow$ (|(xint[3] -- xint[1])*(yint[4] -- yint[2]) -- 

xint[4] - xint[2])*(yint[3] -- yint[1])| /2)    :QUAD AREA

 for i $\leftarrow$ 1 to 4 by 1                     :TRANSLATE+ROTATE ELLIPSE 2

    xint_tr[i] $\leftarrow$ (xint[i] -- H2_TR)*cos ($\varphi $1 -- $\varphi $2)

      + (yint[i] -- K2_TR)*sin ($\varphi $2 -- $\varphi $1)

    yint_tr[i] $\leftarrow$ (xint[i] -- H2_TR)*sin ($\varphi $1 -- $\varphi $2)

      + (yint[i] -- K2_TR)*cos ($\varphi $1 -- $\varphi $2)

    do if ($|xint_tr[i]|>$ A2)             :AVOID INVERSE TRIG ERRORS

       then do if (xint_tr[i] $<$ 0)

               then xint_tr[i] $\leftarrow$ -A2

               else xint_tr[i] $\leftarrow$ A2

    do if (yint_tr[i] $<$ 0) :FIND PARAM ANGLES FOR (xint_tr, yint_tr)

       then $\theta $_tr[i] $\leftarrow$ 2$\pi $ -- arccos (xint_tr[i]/A2)

       else $\theta $_tr[i] $\leftarrow$ arccos (xint_tr[i]/A2)

 xmid $\leftarrow$ A1*cos (($\theta $1 + $\theta $2)/2)

 ymid $\leftarrow$ B1*sin (($\theta $1 + $\theta $2)/2)

 do if (AA*xmid${}^{2}$+BB*xmid*ymid+CC*ymid${}^{2}$+DD*xmid+EE*ymid+FF $<$ 0)

    then area2 = (A1*B1*($\theta $[2] - $\theta $[1])

   - |(xint[1]*yint[2] - xint[2]*yint[1])|)/2

         area3 = (A1*B1*($\theta $[4] - $\theta $[3])

   - |(xint[3]*yint[4] - xint[4]*yint[3])|)/2

         area4 = (A2*B2*($\theta $_tr[3] - $\theta $_tr[2])

   - |(xint\_tr[2]*yint_tr[3] - xint_tr[3]*yint_tr[2])| )/2

         area5 = (A2*B2*($\theta $_tr[1] - $\theta $_tr[4] - twopi))

   - |(xint_tr[4]*yint_tr[1] - xint_tr[1]*yint_tr[4])| /2)

    else area2 = (A1*B1*($\theta $[3] - $\theta $[2])

    - |(xint[2]*yint[3] - xint[3]*yint[2])|)/2

         area3 = (A1*B1*($\theta $[1] - ($\theta $[4] - twopi))

   - |(xint[4]*yint[1] - xint[1]*yint[4])|)/2

         area4 = (A2*B2*($\theta $_tr[2] - $\theta $_tr[1])

   - |(xint_tr[1]*yint_tr[2] - xint_tr[2]*yint_tr[1])|)/2

         area5 = (A2*B2*($\theta $_tr[4] - $\theta $_tr[3])

   - |(xint_tr[3]*yint_tr[4] - xint_tr[4]*yint_tr[3])|)/2

 return (area1+area2+area3+area4+area5, FOUR_INTERSECTION_POINTS)
\end{lstlisting}

\lstset{language=FORTRAN, 
      stringstyle=\color{black},
  keywordstyle=\color{cyan},
caption={The ISTANPT subroutine.  Given an intersection point (\textit{x}, \textit{y}) that satisfies both Ellipse Eq.\ref{eq1} and Ellipse Eq.~\ref{eq2}, the function determines whether the two ellipse curves are tangent at (\textit{x}, \textit{y}), or if the ellipse curves cross at (\textit{x}, \textit{y}).\\},label=code9
   }
\begin{lstlisting}[mathescape][firstnumber=1]
 Code $\leftarrow$ ISTANPT (x,y,A1,B1,AA,BB,CC,DD,EE,FF)

 do if ($|x|>$ A1)                        :AVOID INVERSE TRIG ERRORS

    then do if x $<$ 0

            then x $\leftarrow$ -A1

            else x $\leftarrow$ A1

 do if (y $<$ 0)                    :FIND PARAMETRIC ANGLE FOR (x, y)

    then $\theta $ $\leftarrow$ 2$\pi $ -- arccos (x/A1)

    else $\theta $ $\leftarrow$ arccos (x/A1)

 branch $\leftarrow$ v(x${}^{2}$ + y${}^{2}$)                   :DETERMINE PERTURBATION ANGLE

 do if (branch $<$ 100*EPS)

    then eps_radian $\leftarrow$ 2*EPS

    else eps_radian $\leftarrow$ arcsin (2*EPS/branch)

 x1 $\leftarrow$ A1*cos ($\theta $ + eps_radian)      :CREATE TEST POINTS ON EACH SIDE

 y1 $\leftarrow$ B1*cos ($\theta $ + eps_radian)      :OF THE INPUT POINT (x, y)

 x2 $\leftarrow$ A1*cos ($\theta $ - eps_radian)

 y2 $\leftarrow$ B1*cos ($\theta $ - eps_radian)

 test1 $\leftarrow$ AA*x1${}^{2}$+BB*x1*y1+CC*y1${}^{2}$+DD*x1+EE*y1+FF

 test2 $\leftarrow$ AA*x2${}^{2}$+BB*x2*y2+CC*y2${}^{2}$+DD*x2+EE*y2+FF

 do if (test1*test2 $>$ 0)

     then return TANGENT_POINT

     else return INTERSECTION_POINT
\end{lstlisting}

\section{APPENDIX A}
\label{appA}
\lstset{language=C++, 
      stringstyle=\color{black},
  keywordstyle=\color{cyan},
caption={C-SOURCE CODE FOR ELLIPSE\_SEGMENT\\}
,label=codeappA
   }
\begin{lstlisting}[mathescape][firstnumber=1]
 

 /****************************************************************************

 *

 *  Function: double ellipse_segment

 *

 *  Purpose:  Given the parameters of an ellipse and two points that lie on

 *            the ellipse, this function calculates the ellipse segment area 

 *            between the secant line and the ellipse.  Points are input as

 *            (X1, Y1) and (X2, Y2), and the segment area is defined to be 

 *            between the secant line and the ellipse from the first point

 *            (X1, Y1) to the second point (X2, Y2) in the counter-clockwise

 *            direction.

 *

 *  Reference: Hughes and Chraibi (2011), Calculating Ellipse Overlap Areas

 *

 *  Dependencies: math.h   for calls to trig and absolute value functions  

 *                program_constants.h  error message codes and constants

 *

 *  Inputs:   1. double A      ellipse semi-axis length in x-direction

 *            2. double B      ellipse semi-axis length in y-direction 

 *            3. double X1     x-value of the first point on the ellipse

 *            4. double Y1     y-value of the first point on the ellipse 

 *            5. double X2     x-value of the second point on the ellipse 

 *            6. double Y2     y-value of the second point on the ellipse 

 *

 *  Outputs:  1. int *MessageCode  stores diagnostic information

 *                                 integer codes in program_constants.h

 *

 *  Return:   The value of the ellipse segment area:

 *            -1.0 is returned in case of an error with input data

 *

 ****************************************************************************/

  

 //===========================================================================

 //== INCLUDE ANSI C SYSTEM AND USER-DEFINED HEADER FILES ====================

 //===========================================================================

 #include "program_constants.h"

  

 double ellipse_segment (double A, double B,double X1, double Y1, double X2,

                         double Y2, int *MessageCode)

 {

     double theta1;  //-- parametric angle of the first point

     double theta2;  //-- parametric angle of the second point

     double trsign;  //-- sign of the triangle area

     double pi = 2.0 * asin \eqref{GrindEQ__1_0_};   //-- a maximum-precision value of pi

     double twopi = 2.0 * pi;        //-- a maximum-precision value of 2*pi

  

     //-- Check the data first

     //-- Each of the ellipse axis lengths must be positive

     if (!(A $>$ 0.0) \textbar \textbar  !(B $>$ 0.0))

     {

         (*MessageCode) = ERROR_ELLIPSE_PARAMETERS;

         return -1.0;

     }

  

     //-- Points must be on the ellipse, within EPS, which is defined

     //-- in the header file program_constants.h

     if ( (fabs ((X1*X1)/(A*A) + (Y1*Y1)/(B*B) - 1.0) $>$ EPS) textbar textbar  

          (fabs ((X2*X2)/(A*A) + (Y2*Y2)/(B*B) - 1.0) $>$ EPS) )

     {

         (*MessageCode) = ERROR_POINTS_NOT_ON_ELLIPSE;

         return -1.0;

     }

     

     //-- Avoid inverse trig calculation errors: there could be an error 

     //-- if \textbar X1/A\textbar  $>$ 1.0 or \textbar X2/A\textbar  $>$ 1.0 when calling acos()

     //-- If execution arrives here, then the point is on the ellipse 

     //-- within EPS.  Try to adjust the value of X1 or X2 before giving

     //-- up on the area calculation

     if (fabs (X1)/A $>$ 1.0)

     {

         //-- if execution arrives here, already know that \textbar X1\textbar  $>$ A

         if ((fabs (X1) - A) $>$ EPS)

         {

             //-- if X1 is not close to A or -A, then give up

             (*MessageCode) = ERROR\_INVERSE\_TRIG;

             return -1.0;

         }

         else

         {

             //-- nudge X1 back to A or -A, so acos() will work

             X1 = (X1 $<$ 0) ? -A : A;

         }

     }

  

     if (fabs (X2)/A $>$ 1.0)

     {

         //-- if execution arrives here, already know that \textbar X2\textbar  $>$ A

         if ((fabs (X2) - A) $>$ EPS)

         {

             //-- if X2 is not close to A or -A, then give up

             (*MessageCode) = ERROR_INVERSE_TRIG;

             return -1.0;

         }

         else

         {

             //-- nudge X2 back to A or -A, so acos() will work

             X2 = (X2 $<$ 0) ? -A : A;

         }

     }

  

     //-- Calculate the parametric angles on the ellipse

     //-- The parametric angles depend on the quadrant where each point

     //-- is located.  See Table 1 in the reference.

     if (Y1 $<$ 0.0)    //-- Quadrant III or IV

         theta1 = twopi - acos (X1 / A);

     else             //-- Quadrant I or II      

         theta1 = acos (X1 / A);

         

     if (Y2 $<$ 0.0)    //-- Quadrant III or IV

         theta2 = twopi - acos (X2 / A);

     else             //-- Quadrant I or II      

         theta2 = acos (X2 / A);

     

     //-- need to start the algorithm with theta1 $<$ theta2

     if (theta1 $>$ theta2)

         theta1 -= twopi;

     

     //-- if the integration angle is less than pi, subtract the triangle

     //-- area from the sector, otherwise add the triangle area.

     if ((theta2 - theta1) $>$ pi)

         trsign = 1.0;

     else

         trsign = -1.0;

     

     //-- The ellipse segment is the area between the line and the ellipse, 

     //-- calculated by finding the area of the radial sector minus the area 

     //-- of the triangle created by the center of the ellipse and the two 

     //-- points.  First term is for the ellipse sector; second term is for

     //-- the triangle between the points and the origin.  Area calculation

     //-- is described in the reference.

     (*MessageCode) = NORMAL_TERMINATION;

     return ( 0.5*(A*B*(theta2 - theta1) + trsign*fabs (X1*Y2 - X2*Y1)) );

  

 }

\end{lstlisting}

\section{APPENDIX B}
\label{appB}
\lstset{language=C++, 
      stringstyle=\color{black},
  keywordstyle=\color{cyan},
caption={C-SOURCE CODE FOR ELLIPSE\_LINE\_OVERLAP\\}
,label=codeappB
   }
\begin{lstlisting}[mathescape][firstnumber=1]
 

 /****************************************************************************

 *

 *  Function: double ellipse_line_overlap

 *

 *  Purpose:  Given the parameters of an ellipse and two points on a line, 

 *            this function calculates the area between the two curves.  If 

 *            the line does not cross the ellipse, or if the line is tangent 

 *            to the ellipse, then this function returns an area of 0.0

 *            If the line intersects the ellipse at two points, then the 

 *            function returns the area between the secant line and the 

 *            ellipse.  The line is considered to have a direction from

 *            the first given point (X1,Y1) to the second given point (X2,Y2)

 *            This function determines where the line crosses the ellipse

 *            first, and where it crosses second.  The area returned is 

 *            between the secant line and the ellipse traversed counter-

 *            clockwise from the first intersection point to the second

 *            intersection point.

 *

 *  Reference: Hughes and Chraibi (2011), Calculating Ellipse Overlap Areas

 *

 *  Dependencies: math.h  for calls to trig and absolute value functions  

 *                program_constants.h  error message codes and constants

 *                ellipse_segment.c  core algorithm for ellipse segment area

 *

 *  Inputs:   1. double PHI    CCW rotation angle of the ellipse, radians

 *            2. double A      ellipse semi-axis length in x-direction

 *            3. double B      ellipse semi-axis length in y-direction

 *            4. double H      horizontal offset of ellipse center

 *            5. double K      vertical offset of ellipse center

 *            6. double X1     x-value of the first point on the line

 *            7. double Y1     y-value of the first point on the line 

 *            8. double X2     x-value of the second point on the line 

 *            9. double Y2     y-value of the second point on the line 

 *

 *  Outputs:  1. int *MessageCode  returns diagnostic information

 *                                 integer codes in program_constants.h

 *

 *  Return:   The value of the ellipse segment area:

 *            -1.0 is returned in case of an error with the data or 

 *            calculation

 *            0.0 is returned if the line does not cross the ellipse, or if

 *            the line is tangent to the ellipse

 *

 ****************************************************************************/

  

 //===========================================================================

 //== DEFINE PROGRAM CONSTANTS ===============================================

 //===========================================================================

 #include "program_constants.h"  //-- error message codes and constants

  

 //===========================================================================

 //== DEPENDENT FUNCTIONS ====================================================

 //===========================================================================

 double textbf{ellipse_segment} (double A, double B,double X1, double Y1, double X2,

                         double Y2, int *MessageCode);

  

 double \textbf{ellipse_line_overlap} (double PHI, double A, double B, double H,

                              double K, double X1, double Y1, double X2,

                              double Y2, int *MessageCode)

 \{

     //=======================================================================

     //== DEFINE LOCAL VARIABLES =============================================

     //=======================================================================

     double X10;     //-- Translated, Rotated x-value of the first point 

     double Y10;     //-- Translated, Rotated y-value of the first point

     double X20;     //-- Translated, Rotated x-value of the second point

     double Y20;     //-- Translated, Rotated y-value of the second point

     double cosphi = textbf{cos} (PHI);  //-- store cos(PHI) to avoid multiple calcs

     double sinphi = \textbf{sin} (PHI);  //-- store sin(PHI) to avoid multiple calcs

     double m;       //-- line slope, calculated from input line slope

     double a, b, c;   //-- quadratic equation coefficients a*x\^{}2 + b*x + c

     double discrim;   //-- quadratic equationdiscriminant b\^{}2 - 4*a*c

     double x1, x2;    //-- x-values of intersection points

     double y1, y2;    //-- y-values of intersection points

     double mid_X;     //-- midpoint of the rotated x-values on the line

     double theta1parm;  //-- parametric angle of first point

     double theta2parm;  //-- parametric angle of second point

     double xmidpoint;   //-- x-value midpoint of secant line

     double ymidpoint;   //-- y-value midpoint of secant line

     double root1, root2;  //-- temporary storage variables for roots

     double segment_area;  //-- stores the ellipse segment area

 

     //-- Check the data first

     //-- Each of the ellipse axis lengths must be positive

     if (!(A $>$ 0.0) \textbar \textbar  !(B $>$ 0.0))

     {

         (*MessageCode) = ERROR_ELLIPSE_PARAMETERS;

         return -1.0;

     }

  

     //-- The rotation angle for the ellipse should be between -2pi and 2pi (?)

     if ( (\textbf{fabs} (PHI) $>$ (2.0*pi)) )

         PHI = \textbf{fmod} (PHI, twopi);

     

     //-- For this numerical routine, the ellipse will be translated and

     //-- rotated so that it is centered at the origin and oriented with 

     //-- the coordinate axes.

     //-- Then, the ellipse will have the implicit (polynomial) form of

     //--   x\^{}2/A\^{}2 + y+2/B\^{}2 = 1

     

     //-- For the line, the given points are first translated by the amount

     //-- required to put the ellipse at the origin, e.g., by (-H, -K).  

     //-- Then, the points are rotated by the amount required to orient 

     //-- the ellipse with the coordinate axes, e.g., through the angle -PHI.

     X10 = cosphi*(X1 - H) + sinphi*(Y1 - K);

     Y10 = -sinphi*(X1 - H) + cosphi*(Y1 - K);

     X20 = cosphi*(X2 - H) + sinphi*(Y2 - K);

     Y20 = -sinphi*(X2 - H) + cosphi*(Y2 - K);

     

     //-- To determine if the line and ellipse intersect, solve the two

     //-- equations simultaneously, by substituting y = Y10 + m*(x - X10) 

     //-- and x = X10 + mxy*(y - Y10) into the ellipse equation, 

     //-- which results in two quadratic equations in x.  See the reference

     //-- for derivations of the quadratic coefficients.

     

     //-- If the new line is not close to being vertical, then use the 

     //-- first derivation

     if (\textbf{fabs} (X20 - X10) $>$ EPS)

     {

         //--   ((B\^{}2 + A\^{}2*m\^{}2)/(A\^{}2)) * x\^{}2

         //--   2*(Y10*m - m\^{}2*X10) * x

         //--   (Y10\^{}2 - 2*m*Y10*X10 + m\^{}2*X10\^{}2 - B\^{}2)

         m = (Y20 - Y10)/(X20 - X10);

         a = (B*B + A*A*m*m)/(A*A);

         b = 2.0*(Y10*m - m*m*X10);

         c = (Y10*Y10 - 2.0*m*Y10*X10 + m*m*X10*X10 - B*B);

     }

     //-- If the new line is close to being vertical, then use the 

     //-- second derivation

     else if (\textbf{fabs} (Y20 - Y10) $>$ EPS)

     {

         //--   ((A\^{}2 + B\^{}2*m\^{}2)/(B\^{}2)) * y\^{}2

         //--   2*(X10*m - m\^{}2*Y10) * y

         //--   (X10\^{}2 - 2*m*Y10*X10 + m\^{}2*Y10\^{}2 - A\^{}2)

         m = (X20 - X10)/(Y20 - Y10);

         a = (A*A + B*B*m*m)/(B*B);

         b = 2.0*(X10*m - m*m*Y10);

         c = (X10*X10 - 2.0*m*Y10*X10 + m*m*Y10*Y10 - A*A);

     }

     //-- If the two given points on the line are very close together in 

     //-- both x and y directions, then give up

     else

     {

         (*MessageCode) = ERROR_LINE_POINTS;

         return -1.0;

     }

         

     //-- Once the coefficients for the Quadratic Equation in x are

     //-- known, the roots of the quadratic polynomial will represent 

     //-- the x- or y-values of the points of intersection of the line 

     //-- and the ellipse.  The discriminant can be used to discern   

     //-- which case has occurred for the given inputs:

     //--    1. discr $<$ 0

     //--       Quadratic has complex conjugate roots.

     //--       The line and ellipse do not intersect

     //--    2. discr = 0

     //--       Quadratic has one repeated root

     //--       The line and ellipse intersect at only one point

     //--       i.e., the line is tangent to the ellipse

     //--    3. discr $>$ 0

     //--       Quadratic has two distinct real roots

     //--       The line crosses the ellipse at two points

     discrim = b*b - 4.0*a*c;

     if (discrim $<$ 0.0)

     {

         //-- Line and ellipse do not intersect

         (*MessageCode) = NO_INTERSECTION_POINTS;

         return 0.0;

     }

     else if (discrim $>$ 0.0)

     {

         //-- Two real roots exist, so calculate them

         //-- The larger root is stored in root2

         root1 = (-b - \textbf{sqrt} (discrim)) / (2.0*a);

         root2 = (-b + \textbf{sqrt} (discrim)) / (2.0*a);

     }

     else

     {

         //-- Line is tangent to the ellipse

         (*MessageCode) = LINE_TANGENT_TO_ELLIPSE;

         return 0.0;

     }

  

     //-- decide which roots go into which x or y values

     if (\textbf{fabs} (X20 - X10) $>$ EPS) //-- roots are x-values

     {

         //-- order the points in the same direction as X10 -$>$ X20

         if (X10 $<$ X20)

         {

             x1 = root1;

             x2 = root2;

         }

         else

         {

             x1 = root2;

             x2 = root1;

         }

  

         //-- The y-values can be calculated by substituting the

         //-- x-values into the line equation y = Y10 + m*(x - X10)

         y1 = Y10 + m*(x1 - X10);

         y2 = Y10 + m*(x2 - X10);

     }

     else    //-- roots are y-values

     {

         //-- order the points in the same direction as Y10 -$>$ Y20

         if (Y10 $<$ Y20)

         {

             y1 = root1;

             y2 = root2;

         }

         else

         {

             y1 = root2;

             y2 = root1;

         }

  

         //-- The x-values can be calculated by substituting the

         //-- y-values into the line equation x = X10 + m*(y - Y10)

         x1 = X10 + m*(y1 - Y10);

         x2 = X10 + m*(y2 - Y10);

     }

     

     //-- Arriving here means that two points of intersection have been

     //-- found. Pass the ellipse parameters and intersection points to 

     //-- the ellipse_segment() routine.

     segment_area = \textbf{ellipse_segment} (A, B, x1, y1, x2, y2, MessageCode);

     

     //-- The message code will indicate whether the function encountered

     //-- any errors

     if ((*MessageCode) $<$ 0)

     {

         return -1;

     }

     else

     {

         (*MessageCode) = TWO_INTERSECTION_POINTS;

         return segment_area;

     }

 }
\end{lstlisting}

\section{APPENDIX C}
\label{appC}
\lstset{language=C++, 
      stringstyle=\color{black},
  keywordstyle=\color{cyan},
caption={C-SOURCE CODE FOR ELLIPSE\_ELLIPSE\_OVERLAP\\}
,label=codeappC
   }
\begin{lstlisting}[mathescape][firstnumber=1]
/****************************************************************************

*

*  Function: double ellipse_ellipse_overlap

*

*  Purpose:  Given the parameters of two ellipses, this function calculates

*            the area of overlap between the two curves.  If the ellipses are

*            disjoint, this function returns 0.0; if one ellipse is contained

*            within the other, this function returns the area of the enclosed

*            ellipse; if the ellipses intersect, this function returns the

*            calculated area of overlap.

*

*  Reference: Hughes and Chraibi (2011), Calculating Ellipse Overlap Areas

*

*  Dependencies: math.h  for calls to trig and absolute value functions  

*                program_constants.h  error message codes and constants

*

*  Inputs:   1. double PHI_1  CCW rotation angle of first ellipse, radians

*            2. double A1     semi-axis length in x-direction first ellipse

*            3. double B1     semi-axis length in y-direction first ellipse 

*            4. double H1     horizontal offset of center first ellipse

*            5. double K1     vertical offset of center first ellipse

*            6. double PHI_2  CCW rotation angle of second ellipse, radians

*            7. double A2     semi-axis length in x-direction second ellipse

*            8. double B2     semi-axis length in y-direction second ellipse

*            9. double H2     horizontal offset of center second ellipse 

*           10. double K2     vertical offset of center second ellipse

*

*  Outputs:  1. int *rtnCode  returns diagnostic information integer code

*                                 integer codes in program_constants.h

*

*  Return:   The calculated value of the overlap area

*            -1 is returned in case of an error with the calculation

*             0 is returned if the ellipses are disjoint

*             pi*A*B of smaller ellipse if one ellipse is contained within

*                    the other ellipse

*

****************************************************************************/

 

//===========================================================================

//== DEFINE PROGRAM CONSTANTS ===============================================

//===========================================================================

#include "program_constants.h"  //-- error message codes and constants

 

//===========================================================================

//== DEPENDENT FUNCTIONS ====================================================

//===========================================================================

double nointpts (double A1, double B1, double A2, double B2, double H1, 

                 double K1, double H2_TR, double K2_TR, double AA, double BB, 

                 double CC, double DD, double EE, double FF, int *rtnCode);

 

double twointpts (double xint[], double yint[], double A1, double B1, 

                  double PHI_1, double A2, double B2, double H2_TR, 

                  double K2_TR, double PHI_2, double AA, double BB, 

                  double CC, double DD, double EE, double FF, int *rtnCode);

 

double threeintpts (double xint[], double yint[], double A1, double B1, 

                    double PHI_1, double A2, double B2, double H2_TR, 

                    double K2_TR, double PHI_2, double AA, double BB, 

                    double CC, double DD, double EE, double FF,

                    int *rtnCode);

 

double fourintpts (double xint[], double yint[], double A1, double B1, 

                   double PHI_1, double A2, double B2, double H2_TR, 

                   double K2_TR, double PHI_2, double AA, double BB, 

                   double CC, double DD, double EE, double FF, int *rtnCode);

 

int istanpt (double x, double y, double A1, double B1, double AA, double BB,

             double CC, double DD, double EE, double FF);

 

double ellipse2tr (double x, double y, double AA, double BB, 

                   double CC, double DD, double EE, double FF);

 

//-- functions for solving the quartic equation from Netlib/TOMS

void BIQUADROOTS (double p[], double r[][5]);

void CUBICROOTS (double p[], double r[][5]);

void QUADROOTS (double p[], double r[][5]);

 

//===========================================================================

//== ELLIPSE-ELLIPSE OVERLAP ================================================

//===========================================================================

double ellipse_ellipse_overlap (double PHI_1, double A1, double B1, 

                                double H1, double K1, double PHI_2, 

                                double A2, double B2, double H2, double K2, 

                                int *rtnCode)

{

    //=======================================================================

    //== DEFINE LOCAL VARIABLES =============================================

    //=======================================================================

    int i, j, k, nroots, nychk, nintpts, fnRtnCode;

    double AA, BB, CC, DD, EE, FF, H2_TR, K2_TR, A22, B22, PHI_2R;

    double cosphi, cosphi2, sinphi, sinphi2, cosphisinphi;

    double tmp0, tmp1, tmp2, tmp3;

    double cy[5] = {0.0}, py[5] = {0.0}, r[3][5] = {0.0};

    double x1, x2, y12, y22;

    double ychk[5] = {0.0}, xint[5], yint[5];

    double Area1, Area2, OverlapArea;

 

    //=======================================================================

    //== DATA CHECK =========================================================

    //=======================================================================

    //-- Each of the ellipse axis lengths must be positive

    if ( (!(A1 $>$ 0.0) \textbar \textbar  !(B1 $>$ 0.0)) \textbar \textbar  (!(A2 $>$ 0.0) \textbar \textbar  !(B2 $>$ 0.0)) )

    {

        (*rtnCode) = ERROR_ELLIPSE_PARAMETERS;

        return -1.0;

    }

 

    //-- The rotation angles should be between -2pi and 2pi (?)

    if ( (fabs (PHI_1) $>$ (twopi)) )

        PHI_1 = fmod (PHI_1, twopi);

    if ( (fabs (PHI_2) $>$ (twopi)) )

        PHI_2 = fmod (PHI_2, twopi);

 

    //=======================================================================

    //== DETERMINE THE TWO ELLIPSE EQUATIONS FROM INPUT PARAMETERS ==========

    //=======================================================================

    //-- Finding the points of intersection between two general ellipses

    //-- requires solving a quartic equation.  Before attempting to solve the

    //-- quartic, several quick tests can be used to eliminate some cases

    //-- where the ellipses do not intersect.  Optionally, can whittle away

    //-- at the problem, by addressing the easiest cases first.

 

    //-- Working with the translated+rotated ellipses simplifies the

    //-- calculations.  The ellipses are translated then rotated so that the

    //-- first ellipse is centered at the origin and oriented with the 

    //-- coordinate axes.  Then, the first ellipse will have the implicit 

    //-- (polynomial) form of

    //--   x\^{}2/A1\^{}2 + y+2/B1\^{}2 = 1

    

    //-- For the second ellipse, the center is first translated by the amount

    //-- required to put the first ellipse at the origin, e.g., by (-H1, -K1)  

    //-- Then, the center of the second ellipse is rotated by the amount

    //-- required to orient the first ellipse with the coordinate axes, e.g.,

    //-- through the angle -PHI_1.

    //-- The translated and rotated center point coordinates for the second

    //-- ellipse are found with the rotation matrix, derivations are 

    //-- described in the reference.

    cosphi = cos (PHI_1);

    sinphi = sin (PHI_1);

    H2_TR = (H2 - H1)*cosphi + (K2 - K1)*sinphi;

    K2_TR = (H1 - H2)*sinphi + (K2 - K1)*cosphi;

    PHI_2R = PHI_2 - PHI_1;

    if ( (fabs (PHI_2R) $>$ (twopi)) )

        PHI_2R = fmod (PHI_2R, twopi);

        

    //-- Calculate implicit (Polynomial) coefficients for the second ellipse

    //-- in its translated-by (-H1, -H2) and rotated-by -PHI_1 postion

    //--       AA*x^{}2 + BB*x*y + CC*y^{}2 + DD*x + EE*y + FF = 0

    //-- Formulas derived in the reference

    //-- To speed things up, store multiply-used expressions first 

    cosphi = cos (PHI_2R);

    cosphi2 = cosphi*cosphi;

    sinphi = sin (PHI_2R);

    sinphi2 = sinphi*sinphi;

    cosphisinphi = 2.0*cosphi*sinphi;

    A22 = A2*A2;

    B22 = B2*B2;

    tmp0 = (cosphi*H2_TR + sinphi*K2_TR)/A22;

    tmp1 = (sinphi*H2_TR - cosphi*K2_TR)/B22;

    tmp2 = cosphi*H2_TR + sinphi*K2_TR;

    tmp3 = sinphi*H2_TR - cosphi*K2_TR;

 

    //-- implicit polynomial coefficients for the second ellipse

    AA = cosphi2/A22 + sinphi2/B22;

    BB = cosphisinphi/A22 - cosphisinphi/B22;

    CC = sinphi2/A22 + cosphi2/B22;

    DD = -2.0*cosphi*tmp0 - 2.0*sinphi*tmp1;

    EE = -2.0*sinphi*tmp0 + 2.0*cosphi*tmp1;

    FF = tmp2*tmp2/A22 + tmp3*tmp3/B22 - 1.0;  

    

    //=======================================================================

    //== CREATE AND SOLVE THE QUARTIC EQUATION TO FIND INTERSECTION POINTS ==

    //=======================================================================

    //-- If execution arrives here, the ellipses are at least 'close' to

    //-- intersecting.

    //-- Coefficients for the Quartic Polynomial in y are calculated from

    //-- the two implicit equations.

    //-- Formulas for these coefficients are derived in the reference.

    cy[4] = pow (A1, 4.0)*AA*AA + B1*B1*(A1*A1*(BB*BB - 2.0*AA*CC)

            + B1*B1*CC*CC);

    cy[3] = 2.0*B1*(B1*B1*CC*EE + A1*A1*(BB*DD - AA*EE));

    cy[2] = A1*A1*((B1*B1*(2.0*AA*CC - BB*BB) + DD*DD  - 2.0*AA*FF) 

                   - 2.0*A1*A1*AA*AA) + B1*B1*(2.0*CC*FF + EE*EE);

    cy[1] = 2.0*B1*(A1*A1*(AA*EE - BB*DD) + EE*FF);

    cy[0] = (A1*(A1*AA - DD) + FF)*(A1*(A1*AA + DD) + FF);

 

    //-- Once the coefficients for the Quartic Equation in y are known, the

    //-- roots of the quartic polynomial will represent y-values of the 

    //-- intersection points of the two ellipse curves.

    //-- The quartic sometimes degenerates into a polynomial of lesser 

    //-- degree, so handle all possible cases.

    if (fabs (cy[4]) $>$ 0.0)

    {

        //== QUARTIC COEFFICIENT NONZERO, USE QUARTIC FORMULA ===============

        for (i = 0; i $<$= 3; i++)

            py[4-i] = cy[i]/cy[4];

        py[0] = 1.0;

        

        BIQUADROOTS (py, r);

        nroots = 4;

    }

    else if (fabs (cy[3]) $>$ 0.0)

    {

        //== QUARTIC DEGENERATES TO CUBIC, USE CUBIC FORMULA ================

        for (i = 0; i $<$= 2; i++)

            py[3-i] = cy[i]/cy[3];

        py[0] = 1.0;

 

        CUBICROOTS (py, r);

        nroots = 3;

    }

    else if (fabs (cy[2]) $>$ 0.0)

    {

        //== QUARTIC DEGENERATES TO QUADRATIC, USE QUADRATIC FORMULA ========

        for (i = 0; i $<$= 1; i++)

            py[2-i] = cy[i]/cy[2];

        py[0] = 1.0;

 

        QUADROOTS (py, r);

        nroots = 2;

    }

    else if (fabs (cy[1]) $>$ 0.0)

    {

        //== QUARTIC DEGENERATES TO LINEAR: SOLVE DIRECTLY ==================

        //-- cy[1]*Y + cy[0] = 0

        r[1][1] = (-cy[0]/cy[1]);

        r[2][1] = 0.0;

        nroots = 1;

    }

    else

    {

        //== COMPLETELY DEGENERATE QUARTIC: ELLIPSES IDENTICAL??? ===========

        //-- a completely degenerate quartic, which would seem to

        //-- indicate that the ellipses are identical.  However, some

        //-- configurations lead to a degenerate quartic with no

        //-- points of intersection.

        nroots = 0;

    }

    

    //=======================================================================

    //== CHECK ROOTS OF THE QUARTIC: ARE THEY POINTS OF INTERSECTION? =======

    //=======================================================================

    //-- determine which roots are real, discard any complex roots

    nychk = 0;

    for (i = 1; i $<$= nroots; i++)

    {

        if (fabs (r[2][i]) $<$ EPS)

        {

            nychk++;

            ychk[nychk] = r[1][i]*B1;

        }

    }

    

    //-- sort the real roots by straight insertion

    for (j = 2; j $<$= nychk; j++)

    {

        tmp0 = ychk[j];

        

        for (k = j - 1; k $>$= 1; k--)

        {

            if (ychk[k] $<$= tmp0)

                break;

            

            ychk[k+1] = ychk[k];

        }

        

        ychk[k+1] = tmp0;

    }

 

    //-- determine whether polynomial roots are points of intersection

    //-- for the two ellipses

    nintpts = 0;

    for (i = 1; i $<$= nychk; i++)

    {

        //-- check for multiple roots

        if ((i $>$ 1) \&\& (fabs (ychk[i] - ychk[i-1]) $<$ (EPS/2.0)))

            continue;

 

        //-- check intersection points for ychk[i]

        if (fabs (ychk[i]) $>$ B1)

            x1 = 0.0;

        else

            x1 = A1*sqrt (1.0 - (ychk[i]*ychk[i])/(B1*B1));

        x2 = -x1;

    

        if (fabs(ellipse2tr(x1, ychk[i], AA, BB, CC, DD, EE, FF)) $<$ EPS/2.0)

        {

            nintpts++;

            if (nintpts $>$ 4)

            {

                (*rtnCode) = ERROR_INTERSECTION_PTS;

                return -1.0;

            }

            xint[nintpts] = x1;

            yint[nintpts] = ychk[i];

        }

 

        if ((fabs(ellipse2tr(x2, ychk[i], AA, BB, CC, DD, EE, FF)) $<$ EPS/2.0)

            \&\& (fabs (x2 - x1) $>$ EPS/2.0))

        {

            nintpts++;

            if (nintpts $>$ 4)

            {

                (*rtnCode) = ERROR_INTERSECTION_PTS;

                return -1.0;

            }

            xint[nintpts] = x2;

            yint[nintpts] = ychk[i];

        }

    }

 

    //=======================================================================

    //== HANDLE ALL CASES FOR THE NUMBER OF INTERSCTION POINTS ==============

    //=======================================================================

    switch (nintpts)

    {

        case 0:

        case 1:

            OverlapArea = nointpts (A1, B1, A2, B2, H1, K1, H2_TR, K2_TR, AA,

                                    BB, CC, DD, EE, FF, rtnCode);

            return OverlapArea;

            

        case 2:

            //-- when there are two intersection points, it is possible for

            //-- them to both be tangents, in which case one of the ellipses

            //-- is fully contained within the other.  Check the points for

            //-- tangents; if one of the points is a tangent, then the other

            //-- must be as well, otherwise there would be more than 2 

            //-- intersection points.

            fnRtnCode = istanpt (xint[1], yint[1], A1, B1, AA, BB, CC, DD, 

                                 EE, FF);

 

            if (fnRtnCode == TANGENT_POINT)

                OverlapArea = nointpts (A1, B1, A2, B2, H1, K1, H2_TR, K2_TR,

                                        AA, BB, CC, DD, EE, FF, rtnCode);

            else

                OverlapArea = twointpts (xint, yint, A1, B1, PHI_1, A2, B2,

                                         H2_TR, K2_TR, PHI_2, AA, BB, CC, DD,

                                         EE, FF, rtnCode);

            return OverlapArea;

            

        case 3:

            //-- when there are three intersection points, one and only one

            //-- of the points must be a tangent point.

            OverlapArea = threeintpts (xint, yint,  A1, B1, PHI_1, A2, B2,

                                       H2_TR, K2_TR, PHI_2, AA, BB, CC, DD, 

                                       EE, FF, rtnCode);

            return OverlapArea;

            

        case 4:

            //-- four intersections points has only one case.

            OverlapArea = fourintpts (xint, yint,  A1, B1, PHI_1, A2, B2,

                                      H2_TR, K2_TR, PHI_2, AA, BB, CC, DD, 

                                      EE, FF, rtnCode);

            return OverlapArea;

        

        default:

            //-- should never get here (but get compiler warning for missing

            //-- return value if this line is omitted)

            (*rtnCode) = ERROR_INTERSECTION_PTS;

            return -1.0;

    }

}

 

double ellipse2tr (double x, double y, double AA, double BB, 

                   double CC, double DD, double EE, double FF)

{

    return (AA*x*x + BB*x*y + CC*y*y + DD*x + EE*y + FF);

}

 

double nointpts (double A1, double B1, double A2, double B2, double H1, 

                 double K1, double H2_TR, double K2_TR, double AA, double BB,

                 double CC, double DD, double EE, double FF, int *rtnCode)

{

    //-- The relative size of the two ellipses can be found from the axis

    //-- lengths 

    double relsize = (A1*B1) - (A2*B2);

    

    if (relsize $>$ 0.0)

    {

        //-- First Ellipse is larger than second ellipse.

        //-- If second ellipse center (H2_TR, K2_TR) is inside

        //-- first ellipse, then ellipse 2 is completely inside 

        //-- ellipse 1. Otherwise, the ellipses are disjoint.

        if ( ((H2_TR*H2_TR) / (A1*A1) 

            + (K2_TR*K2_TR) / (B1*B1)) $<$ 1.0 )

        {

            (*rtnCode) = ELLIPSE2_INSIDE_ELLIPSE1;

            return (pi*A2*B2);

        }

        else

        {

            (*rtnCode) = DISJOINT_ELLIPSES;

            return 0.0;

        }

    }

    else if (relsize $<$ 0.0)

    {

        //-- Second Ellipse is larger than first ellipse

        //-- If first ellipse center (0, 0) is inside the

        //-- second ellipse, then ellipse 1 is completely inside

        //-- ellipse 2. Otherwise, the ellipses are disjoint

        //--   AA*x^{}2 + BB*x*y + CC*y\^{}2 + DD*x + EE*y + FF = 0

        if (FF $<$ 0.0)

        {

            (*rtnCode) = ELLIPSE1_INSIDE_ELLIPSE2;

            return (pi*A1*B1);

        }

        else

        {

            (*rtnCode) = DISJOINT_ELLIPSES;

            return 0.0;

        }

    }

    else

    {

        //-- If execution arrives here, the relative sizes are identical.

        //-- Are the ellipses the same?  Check the parameters to see.

        if ((H1 == H2_TR) \&\& (K1 == K2_TR))

        {

            (*rtnCode) = ELLIPSES_ARE_IDENTICAL;

            return (pi*A1*B1);

        }

        else

        {

            //-- should never get here, so return error

            (*rtnCode) = ERROR_CALCULATIONS;

            return -1.0;

        }

    }//-- end if (relsize $>$ 0.0)

}

 

//-- two distinct intersection points (x1, y1) and (x2, y2) find overlap area

double twointpts (double x[], double y[], double A1, double B1, double PHI_1, 

                  double A2, double B2, double H2_TR, double K2_TR, 

                  double PHI_2, double AA, double BB, double CC, double DD, 

                  double EE, double FF, int *rtnCode)

{

    double area1, area2;

    double xmid, ymid, xmid_rt, ymid_rt;

    double theta1, theta2;

    double tmp, trsign;

    double x1_tr, y1_tr, x2_tr, y2_tr;

    double discr;

    double cosphi, sinphi;

 

    //-- if execution arrives here, the intersection points are not

    //-- tangents.

    

    //-- determine which direction to integrate in the ellipse_segment

    //-- routine for each ellipse.

 

    //-- find the parametric angles for each point on ellipse 1

    if (fabs (x[1]) $>$ A1)

        x[1] = (x[1] $<$ 0) ? -A1 : A1;

    if (y[1] $<$ 0.0)      //-- Quadrant III or IV

        theta1 = twopi - acos (x[1] / A1);

    else             //-- Quadrant I or II      

        theta1 = acos (x[1] / A1);

        

    if (fabs (x[2]) $>$ A1)

        x[2] = (x[2] $<$ 0) ? -A1 : A1;

    if (y[2] $<$ 0.0)      //-- Quadrant III or IV

        theta2 = twopi - acos (x[2] / A1);

    else             //-- Quadrant I or II      

        theta2 = acos (x[2] / A1);

 

    //-- logic is for proceeding counterclockwise from theta1 to theta2

    if (theta1 $>$ theta2)

    {

        tmp = theta1;

        theta1 = theta2;

        theta2 = tmp;

    }

 

    //-- find a point on the first ellipse that is different than the two

    //-- intersection points.

    xmid = A1*cos ((theta1 + theta2)/2.0);  

    ymid = B1*sin ((theta1 + theta2)/2.0);  

    

    //-- the point (xmid, ymid) is on the first ellipse 'between' the two

    //-- intersection points (x[1], y[1]) and (x[2], y[2]) when travelling 

    //-- counter- clockwise from (x[1], y[1]) to (x[2], y[2]).  If the point

    //-- (xmid, ymid) is inside the second ellipse, then the desired segment

    //-- of ellipse 1 contains the point (xmid, ymid), so integrate 

    //-- counterclockwise from (x[1], y[1]) to (x[2], y[2]).  Otherwise, 

    //-- integrate counterclockwise from (x[2], y[2]) to (x[1], y[1])

    if (ellipse2tr (xmid, ymid, AA, BB, CC, DD, EE, FF) $>$ 0.0)

    {

        tmp = theta1;

        theta1 = theta2;

        theta2 = tmp;

    }

 

    //-- here is the ellipse segment routine for the first ellipse

    if (theta1 $>$ theta2)

        theta1 -= twopi;

    if ((theta2 - theta1) $>$ pi)

        trsign = 1.0;

    else

        trsign = -1.0;

    area1 = 0.5*(A1*B1*(theta2 - theta1) 

            + trsign*fabs (x[1]*y[2] - x[2]*y[1])); 

    

    //-- find ellipse 2 segment area.  The ellipse segment routine

    //-- needs an ellipse that is centered at the origin and oriented

    //-- with the coordinate axes.  The intersection points (x[1], y[1]) and

    //-- (x[2], y[2]) are found with both ellipses translated and rotated by

    //-- (-H1, -K1) and -PHI_1.  Further translate and rotate the points

    //-- to put the second ellipse at the origin and oriented with the

    //-- coordinate axes.  The translation is (-H2_TR, -K2_TR), and the

    //-- rotation is -(PHI_2 - PHI_1) = PHI_1 - PHI_2

    cosphi = cos (PHI_1 - PHI_2);

    sinphi = sin (PHI_1 - PHI_2);

    x1_tr = (x[1] - H2_TR)*cosphi + (y[1] - K2_TR)*-sinphi;

    y1_tr = (x[1] - H2_TR)*sinphi + (y[1] - K2_TR)*cosphi;

    x2_tr = (x[2] - H2_TR)*cosphi + (y[2] - K2_TR)*-sinphi;

    y2_tr = (x[2] - H2_TR)*sinphi + (y[2] - K2_TR)*cosphi;

    

    //-- determine which branch of the ellipse to integrate by finding a

    //-- point on the second ellipse, and asking whether it is inside the

    //-- first ellipse (in their once-translated+rotated positions)

    //-- find the parametric angles for each point on ellipse 1

    if (fabs (x1_tr) $>$ A2)

        x1_tr = (x1_tr $<$ 0) ? -A2 : A2;

    if (y1_tr $<$ 0.0)     //-- Quadrant III or IV

        theta1 = twopi - acos (x1_tr/A2);

    else             //-- Quadrant I or II      

        theta1 = acos (x1_tr/A2);

        

    if (fabs (x2_tr) $>$ A2)

        x2_tr = (x2_tr $<$ 0) ? -A2 : A2;

    if (y2_tr $<$ 0.0)     //-- Quadrant III or IV

        theta2 = twopi - acos (x2_tr/A2);

    else             //-- Quadrant I or II      

        theta2 = acos (x2_tr/A2);

 

    //-- logic is for proceeding counterclockwise from theta1 to theta2

    if (theta1 $>$ theta2)

    {

        tmp = theta1;

        theta1 = theta2;

        theta2 = tmp;

    }

 

    //-- find a point on the second ellipse that is different than the two

    //-- intersection points.

    xmid = A2*cos ((theta1 + theta2)/2.0);  

    ymid = B2*sin ((theta1 + theta2)/2.0);

    

    //-- translate the point back to the second ellipse in its once-

    //-- translated+rotated position

    cosphi = cos (PHI_2 - PHI_1);

    sinphi = sin (PHI_2 - PHI_1);

    xmid_rt = xmid*cosphi + ymid*-sinphi + H2_TR;

    ymid_rt = xmid*sinphi + ymid*cosphi + K2_TR;

 

    //-- the point (xmid_rt, ymid_rt) is on the second ellipse 'between' the

    //-- intersection points (x[1], y[1]) and (x[2], y[2]) when travelling

    //-- counterclockwise from (x[1], y[1]) to (x[2], y[2]).  If the point

    //-- (xmid_rt, ymid_rt) is inside the first ellipse, then the desired 

    //-- segment of ellipse 2 contains the point (xmid_rt, ymid_rt), so 

    //-- integrate counterclockwise from (x[1], y[1]) to (x[2], y[2]).  

    //-- Otherwise, integrate counterclockwise from (x[2], y[2]) to 

    //-- (x[1], y[1])

    if (((xmid_rt*xmid_rt)/(A1*A1) + (ymid_rt*ymid_rt)/(B1*B1)) $>$ 1.0)

    {

        tmp = theta1;

        theta1 = theta2;

        theta2 = tmp;

    }

 

    //-- here is the ellipse segment routine for the second ellipse

    if (theta1 $>$ theta2)

        theta1 -= twopi;

    if ((theta2 - theta1) $>$ pi)

        trsign = 1.0;

    else

        trsign = -1.0;

    area2 = 0.5*(A2*B2*(theta2 - theta1) 

            + trsign*fabs (x1_tr*y2_tr - x2_tr*y1_tr)); 

    

    (*rtnCode) = TWO_INTERSECTION_POINTS;

    return area1 + area2;

}

 

//-- three distinct intersection points, must have two intersections

//-- and one tangent, which is the only possibility

double threeintpts (double xint[], double yint[], double A1, double B1, 

                    double PHI_1, double A2, double B2, double H2_TR, 

                    double K2_TR, double PHI_2, double AA, double BB, 

                    double CC, double DD, double EE, double FF,

                    int *rtnCode)

{

    int i, tanpts, tanindex, fnRtn;

    double OverlapArea;

 

    //-- need to determine which point is a tangent, and which two points

    //-- are intersections

    tanpts = 0;

    for (i = 1; i $<$= 3; i++)

    {

        fnRtn = istanpt (xint[i], yint[i], A1, B1, AA, BB, CC, DD, EE, FF);

 

        if (fnRtn == TANGENT_POINT)

        {

            tanpts++;

            tanindex = i;

        }

    }

    

    //-- there MUST be 2 intersection points and only one tangent

    if (tanpts != 1)

    {

        //-- should never get here unless there is a problem discerning

        //-- whether or not a point is a tangent or intersection

        (*rtnCode) = ERROR_INTERSECTION_PTS;

        return -1.0;

    }

    

    //-- store the two interesection points into (x[1], y[1]) and 

    //-- (x[2], y[2])

    switch (tanindex)

    {

        case 1:

            xint[1] = xint[3];

            yint[1] = yint[3];

            break;

 

        case 2:

            xint[2] = xint[3];

            yint[2] = yint[3];

            break;

 

        case 3:

            //-- intersection points are already in the right places

            break;

    }

 

    OverlapArea = twointpts (xint, yint, A1, B1, PHI_1, A2, B2, H2_TR, K2_TR,

                             PHI_2, AA, BB, CC, DD, EE, FF, rtnCode);

    (*rtnCode) = THREE_INTERSECTION_POINTS;

    return OverlapArea;

}

 

//-- four intersection points

double fourintpts (double xint[], double yint[], double A1, double B1, 

                   double PHI_1, double A2, double B2, double H2_TR, 

                   double K2_TR, double PHI_2, double AA, double BB, 

                   double CC, double DD, double EE, double FF, int *rtnCode)

{

    int i, j, k;

    double xmid, ymid, xint_tr[5], yint_tr[5], OverlapArea;

    double theta[5], theta_tr[5], cosphi, sinphi, tmp0, tmp1, tmp2;

    double area1, area2, area3, area4, area5;

    

    //-- only one case, which involves two segments from each ellipse, plus

    //-- two triangles.

    //-- get the parametric angles along the first ellipse for each of the

    //-- intersection points

    for (i = 1; i $<$= 4; i++)

    {

        if (fabs (xint[i]) $>$ A1)

            xint[i] = (xint[i] $<$ 0) ? -A1 : A1;

        if (yint[i] $<$ 0.0)   //-- Quadrant III or IV

            theta[i] = twopi - acos (xint[i] / A1);

        else             //-- Quadrant I or II      

            theta[i] = acos (xint[i] / A1);

    }

        

    //-- sort the angles by straight insertion, and put the points in 

    //-- counter-clockwise order

    for (j = 2; j $<$= 4; j++)

    {

        tmp0 = theta[j];

        tmp1 = xint[j];

        tmp2 = yint[j];

        

        for (k = j - 1; k $>$= 1; k--)

        {

            if (theta[k] $<$= tmp0)

                break;

            

            theta[k+1] = theta[k];

            xint[k+1] = xint[k];

            yint[k+1] = yint[k];

        }

        

        theta[k+1] = tmp0;

        xint[k+1] = tmp1;

        yint[k+1] = tmp2;

    }

    

    //-- find the area of the interior quadrilateral

    area1 = 0.5*fabs ((xint[3] - xint[1])*(yint[4] - yint[2])

                     - (xint[4] - xint[2])*(yint[3] - yint[1]));

 

    //-- the intersection points lie on the second ellipse in its once

    //-- translated+rotated position.  The segment algorithm is implemented

    //-- for an ellipse that is centered at the origin, and oriented with

    //-- the coordinate axes; so, in order to use the segment algorithm

    //-- with the second ellipse, the intersection points must be further

    //-- translated+rotated by amounts that put the second ellipse centered

    //-- at the origin and oriented with the coordinate axes.

    cosphi = cos (PHI_1 - PHI_2);

    sinphi = sin (PHI_1 - PHI_2);

    for (i = 1; i $<$= 4; i++)

    {

        xint_tr[i] = (xint[i] - H2_TR)*cosphi + (yint[i] - K2_TR)*-sinphi;

        yint_tr[i] = (xint[i] - H2_TR)*sinphi + (yint[i] - K2_TR)*cosphi;

        

        if (fabs (xint_tr[i]) $>$ A2)

            xint_tr[i] = (xint_tr[i] $<$ 0) ? -A2 : A2;

        if (yint_tr[i] $<$ 0.0)    //-- Quadrant III or IV

            theta_tr[i] = twopi - acos (xint_tr[i]/A2);

        else             //-- Quadrant I or II      

            theta_tr[i] = acos (xint_tr[i]/A2);

    }

 

    //-- get the area of the two segments on ellipse 1

    xmid = A1*cos ((theta[1] + theta[2])/2.0);  

    ymid = B1*sin ((theta[1] + theta[2])/2.0);

    

    //-- the point (xmid, ymid) is on the first ellipse 'between' the two

    //-- sorted intersection points (xint[1], yint[1]) and (xint[2], yint[2])

    //-- when travelling counter- clockwise from (xint[1], yint[1]) to 

    //-- (xint[2], yint[2]).  If the point (xmid, ymid) is inside the second 

    //-- ellipse, then one desired segment of ellipse 1 contains the point 

    //-- (xmid, ymid), so integrate counterclockwise from (xint[1], yint[1])

    //-- to (xint[2], yint[2]) for the first segment, and from 

    //-- (xint[3], yint[3] to (xint[4], yint[4]) for the second segment.

    if (ellipse2tr (xmid, ymid, AA, BB, CC, DD, EE, FF) $<$ 0.0)

    {

        area2 = 0.5*(A1*B1*(theta[2] - theta[1])

                - fabs (xint[1]*yint[2] - xint[2]*yint[1]));

        area3 = 0.5*(A1*B1*(theta[4] - theta[3])

                - fabs (xint[3]*yint[4] - xint[4]*yint[3]));

        area4 = 0.5*(A2*B2*(theta_tr[3] - theta_tr[2])

                - fabs (xint_tr[2]*yint_tr[3] - xint_tr[3]*yint_tr[2]));

        area5 = 0.5*(A2*B2*(theta_tr[1] - (theta_tr[4] - twopi))

                - fabs (xint_tr[4]*yint_tr[1] - xint_tr[1]*yint_tr[4]));

    }

    else

    {

        area2 = 0.5*(A1*B1*(theta[3] - theta[2])

                - fabs (xint[2]*yint[3] - xint[3]*yint[2]));

        area3 = 0.5*(A1*B1*(theta[1] - (theta[4] - twopi))

                - fabs (xint[4]*yint[1] - xint[1]*yint[4]));

        area4 = 0.5*(A2*B2*(theta[2] - theta[1])

                - fabs (xint_tr[1]*yint_tr[2] - xint_tr[2]*yint_tr[1]));

        area5 = 0.5*(A2*B2*(theta[4] - theta[3])

                - fabs (xint_tr[3]*yint_tr[4] - xint_tr[4]*yint_tr[3]));

    }

 

    OverlapArea = area1 + area2 + area3 + area4 + area5;

    (*rtnCode) = FOUR_INTERSECTION_POINTS;

    return OverlapArea;

}

 

//-- check whether an intersection point is a tangent or a cross-point

int istanpt (double x, double y, double A1, double B1, double AA, double BB,

             double CC, double DD, double EE, double FF)

{

    double x1, y1, x2, y2, theta, test1, test2, branch, eps_radian;

 

    //-- Avoid inverse trig calculation errors: there could be an error 

    //-- if \textbar x1/A\textbar  $>$ 1.0 when calling acos().  If execution arrives here, 

    //-- then the point is on the ellipse within EPS.

    if (fabs (x) $>$ A1)

        x = (x $<$ 0) ? -A1 : A1;

 

    //-- Calculate the parametric angle on the ellipse for (x, y)

    //-- The parametric angles depend on the quadrant where each point

    //-- is located.  See Table 1 in the reference.

    if (y $<$ 0.0)     //-- Quadrant III or IV

        theta = twopi - acos (x / A1);

    else             //-- Quadrant I or II      

        theta = acos (x / A1);

 

    //-- determine the distance from the origin to the point (x, y)

    branch = sqrt (x*x + y*y);

 

    //-- use the distance to find a small angle, such that the distance

    //-- along ellipse 1 is approximately 2*EPS

    if (branch $<$ 100.0*EPS)

        eps_radian = 2.0*EPS;

    else

        eps_radian = asin (2.0*EPS/branch);

 

    //-- determine two points that are on each side of (x, y) and lie on

    //-- the first ellipse

    x1 = A1*cos (theta + eps_radian);

    y1 = B1*sin (theta + eps_radian);

    x2 = A1*cos (theta - eps_radian);

    y2 = B1*sin (theta - eps_radian);

    

    //-- evaluate the two adjacent points in the second ellipse equation

    test1 = ellipse2tr (x1, y1, AA, BB, CC, DD, EE, FF);

    test2 = ellipse2tr (x2, y2, AA, BB, CC, DD, EE, FF);

 

    //-- if the ellipses are tangent at the intersection point, then

    //-- points on both sides will either both be inside ellipse 1, or

    //-- they will both be outside ellipse 1

    if ((test1*test2) $>$ 0.0)

        return TANGENT_POINT;

    else

        return INTERSECTION_POINT;

}

 

//===========================================================================

//-- CACM Algorithm 326: Roots of low order polynomials.

//-- Nonweiler, Terence R.F., CACM Algorithm 326: Roots of low order 

//-- polynomials, Communications of the ACM, vol. 11 no. 4, pages 

//-- 269-270 (1968). Translated into c and programmed by M. Dow, ANUSF,

//-- Australian National University, Canberra, Australia.

//-- Accessed at http://www.netlib.org/toms/326.

//-- Modified to void functions, integers replaced with floating point

//-- where appropriate, some other slight modifications for readability

//-- and debugging ease.

//===========================================================================

void QUADROOTS (double p[], double r[][5])

{

    /*

    Array r[3][5]  p[5]

    Roots of poly p[0]*x^{}2 + p[1]*x + p[2]=0

    x=r[1][k] + i r[2][k]  k=1,2

    */

    double b,c,d;

    b=-p[1]/(2.0*p[0]);

    c=p[2]/p[0];

    d=b*b-c;

    if(d$>$=0.0)

    {

        if(b$>$0.0) 

            b=(r[1][2]=(sqrt(d)+b));

        else    

            b=(r[1][2]=(-sqrt(d)+b));

        r[1][1]=c/b; 

        r[2][1]=(r[2][2]=0.0);

    }

    else

    {

        d=(r[2][1]=sqrt(-d)); 

        r[2][2]=-d;

        r[1][1]=(r[1][2]=b);

    }

    return;

}

 

void CUBICROOTS(double p[], double r[][5])

{

    /*

    Array r[3][5]  p[5]

    Roots of poly p[0]*x\^{}3 + p[1]*x\^{}2 + p[2]*x + p[3] = 0

    x=r[1][k] + i r[2][k]  k=1,...,3

    Assumes 0$<$arctan(x)$<$pi/2 for x$>$0

    */

    double s,t,b,c,d;

    int k;

    if(p[0]!=1.0)

    {

        for(k=1;k$<$4;k++) 

            p[k]=p[k]/p[0]; 

        p[0]=1.0;

    }

    s=p[1]/3.0; 

    t=s*p[1];

    b=0.5*(s*(t/1.5-p[2])+p[3]);

    t=(t-p[2])/3.0;

    c=t*t*t; 

    d=b*b-c;

    if(d$>$=0.0)

    {

        d=pow((sqrt(d)+fabs(b)),1.0/3.0);

        if(d!=0.0)

        {

            if(b$>$0.0) 

                b=-d;

            else 

                b=d;

            c=t/b;

        }

        d=r[2][2]=sqrt\eqref{GrindEQ__0_75_}*(b-c); 

        b=b+c;

        c=r[1][2]=-0.5*b-s;

        if((b$>$0.0 \&\& s$<$=0.0) \textbar \textbar  (b$<$0.0 \&\& s$>$0.0))

        {

            r[1][1]=c; 

            r[2][1]=-d; 

            r[1][3]=b-s;

            r[2][3]=0.0;

        }

        else

        {

            r[1][1]=b-s; 

            r[2][1]=0.0; 

            r[1][3]=c;

            r[2][3]=-d;

        }

    }  /* end 2 equal or complex roots */

    else

    {

        if(b==0.0)

            d=atan\eqref{GrindEQ__1_0_}/1.5;

        else

            d=atan(sqrt(-d)/fabs(b))/3.0;

        if(b$<$0.0)

            b=2.0*sqrt(t);

        else

            b=-2.0*sqrt(t);

        c=cos(d)*b; 

        t=-sqrt\eqref{GrindEQ__0_75_}*sin(d)*b-0.5*c;

        d=-t-c-s; 

        c=c-s; 

        t=t-s;

        if(fabs(c)$>$fabs(t))

        {

            r[1][3]=c;

        }

        else

        {

            r[1][3]=t; 

            t=c;

        }

        if(fabs(d)$>$fabs(t))

        {

            r[1][2]=d;

        }

        else

        {

            r[1][2]=t; 

            t=d;

        }

        r[1][1]=t;

        for(k=1;k$<$4;k++) 

            r[2][k]=0.0;

    }

    return;

}

 

void BIQUADROOTS(double p[],double r[][5])

{

    /*

    Array r[3][5]  p[5]

    Roots of poly p[0]*x\^{}4 + p[1]*x\^{}3 + p[2]*x\^{}2 + p[3]*x + p[4] = 0

    x=r[1][k] + i r[2][k]  k=1,...,4

    */

    double a,b,c,d,e;

    int k,j;

    if(p[0] != 1.0)

    {

        for(k=1;k$<$5;k++) 

            p[k]=p[k]/p[0];

        p[0]=1.0;

    }

    e=0.25*p[1];

    b=2.0*e;

    c=b*b;

    d=0.75*c;

    b=p[3]+b*(c-p[2]);

    a=p[2]-d;

    c=p[4]+e*(e*a-p[3]);

    a=a-d;

    p[1]=0.5*a;

    p[2]=(p[1]*p[1]-c)*0.25;

    p[3]=b*b/(-64.0);

    if(p[3]$<$0.0)

    {

        CUBICROOTS(p,r);

        for(k=1;k$<$4;k++)

        {

            if(r[2][k]==0.0 \&\& r[1][k]$>$0.0)

            {

                d=r[1][k]*4.0; 

                a=a+d;

                if(a$>$=0.0 \&\& b$>$=0.0)

                    p[1]=sqrt(d);

                else if(a$<$=0.0 \&\& b$<$=0.0)

                    p[1]=sqrt(d);

                else 

                    p[1]=-sqrt(d);

                b=0.5*(a+b/p[1]);

                goto QUAD;

            }

        }

    }

    if(p[2]$<$0.0)

    {

        b=sqrt(c); 

        d=b+b-a;

        p[1]=0.0; 

        if(d$>$0.0) 

            p[1]=sqrt(d);

    }

    else

    {

        if(p[1]$>$0.0)

            b=sqrt(p[2])*2.0+p[1];

        else

            b=-sqrt(p[2])*2.0+p[1];

        if(b!=0.0)

        {

            p[1]=0.0;

        }

        else

        {

            for(k=1;k$<$5;k++)

            {

                r[1][k]=-e;

                r[2][k]=0.0;

            }

            goto END;

        }

    }

QUAD:

    p[2]=c/b; 

    QUADROOTS(p,r);

    for(k=1;k$<$3;k++)

        for(j=1;j$<$3;j++) 

            r[j][k+2]=r[j][k];

    p[1]=-p[1]; 

    p[2]=b; 

    QUADROOTS(p,r);

    for(k=1;k$<$5;k++) 

        r[1][k]=r[1][k]-e;

END:

    return;

}

\end{lstlisting}

\section{APPENDIX D}
\label{appD}
\lstset{language=C++, 
      stringstyle=\color{black},
  keywordstyle=\color{cyan},
caption={C-SOURCE CODE FOR UTILITY FUNCTIONS\\}
,label=codeappD
   }
\begin{lstlisting}[mathescape][firstnumber=1]
program\_constants.h:



//===========================================================================

//== INCLUDE ANSI C SYSTEM HEADER FILES =====================================

//===========================================================================

#include $<$math.h$>$   //-- for calls to trig, sqrt and power functions

 

//==========================================================================

//== DEFINE PROGRAM CONSTANTS ==============================================

//==========================================================================

#define NORMAL_TERMINATION                    0

#define NO_INTERSECTION_POINTS              100

#define ONE_INTERSECTION_POINT              101

#define LINE_TANGENT_TO_ELLIPSE             102

#define DISJOINT_ELLIPSES                   103

#define ELLIPSE2_OUTSIDETANGENT_ELLIPSE1    104

#define ELLIPSE2_INSIDETANGENT_ELLIPSE1     105

#define ELLIPSES_INTERSECT                  106

#define TWO_INTERSECTION_POINTS             107

#define THREE_INTERSECTION_POINTS           108

#define FOUR_INTERSECTION_POINTS            109

#define ELLIPSE1_INSIDE_ELLIPSE2            110

#define ELLIPSE2_INSIDE_ELLIPSE1            111

#define ELLIPSES_ARE_IDENTICAL              112

#define INTERSECTION_POINT                  113

#define TANGENT_POINT                       114

 

#define ERROR_ELLIPSE_PARAMETERS           -100

#define ERROR_DEGENERATE_ELLIPSE           -101

#define ERROR_POINTS_NOT_ON_ELLIPSE        -102

#define ERROR_INVERSE_TRIG                 -103

#define ERROR_LINE_POINTS                  -104

#define ERROR_QUARTIC_CASE                 -105

#define ERROR_POLYNOMIAL_DEGREE            -107

#define ERROR_POLYNOMIAL_ROOTS             -108

#define ERROR_INTERSECTION_PTS             -109

#define ERROR_CALCULATIONS                 -112

 

#define EPS                            +1.0E-07

#define pi     (2.0*asin (1.0)) //-- a maximum-precision value of pi

#define twopi  (2.0*pi)         //-- a maximum-precision value of 2*pi







call_es.c:



#include $<$stdio.h$>$

#include $<$math.h$>$

#include "program_constants.h"

double ellipse_segment (double A, double B, double X1, double Y1, double X2,

                        double Y2, int *MessageCode);

 

int main (int argc, char ** argv) 

{

   double A, B;

   double X1, Y1;

   double X2, Y2;

   double area1, area2;

   double pi = 2.0 * asin eqref{GrindEQ__1_0_};    //-- a maximum-precision value of pi

   int rtn;

   char msg[1024];

   printf ("Calling ellipse_segment.ctextbackslash n");

   

   //-- case shown in Fig. 1

   A = 4.;

   B = 2.;

   X1 = 4./sqrt (5.);

   Y1 = 4./sqrt (5.);

   X2 = -3.;

   Y2 = -sqrt (7.)/2.;

 

   area1 = ellipse_segment (A, B, X1, Y1, X2, Y2, &rtn);

   sprintf (msg,"Fig 1: segment area = %15.8f, return_value = %d\textbackslash n", area1, rtn);

   printf (msg);

   

   //-- case shown in Fig. 2

   A = 4.;

   B = 2.;

   X1 = -3.;

   Y1 = -sqrt (7.)/2.;

   X2 = 4./sqrt (5.);

   Y2 = 4./sqrt (5.);

 

   area2 = ellipse_segment (A, B, X1, Y1, X2, Y2, &rtn);

   sprintf (msg,"Fig 2: segment area = %15.8f, return_value = %dtextbackslash n", area2, rtn);

   printf (msg);

 

   sprintf (msg,"sum of ellipse segments = %15.8ftextbackslash n", area1 + area2);

   printf (msg);

   sprintf (msg,"total ellipse area by pi*a*b = %15.8ftextbackslash n", pi*A*B);

   printf (msg);

 

   return rtn; 

}





call_el.c:



#include $<$stdio.h$>$

#include $<$math.h$>$

#include "program_constants.h"

double \textbf{ellipse_segment} (double A, double B, double X1, double Y1, double X2,

                        double Y2, int *MessageCode);

 

double \textbf{ellipse_line_overlap} (double PHI, double A, double B, double H,

                             double K, double X1, double Y1, double X2,

                             double Y2, int *MessageCode);

                             

int \textbf{main} (int argc, char ** argv) 

{

   double A, B;

   double H, K, PHI;

   double X1, Y1;

   double X2, Y2;

   double area1, area2;

   double pi = 2.0 * \textbf{asin} \eqref{GrindEQ__1_0_};    //-- a maximum-precision value of pi

   int rtn;

   char msg[1024];

   \textbf{printf} ("Calling ellipse_line_overlap.c\textbackslash n");

   

   //-- case shown in Fig. 4

   A = 4.;

   B = 2.;

   H = -6;

   K = 3;

   PHI = 3.*pi/8.0;

   X1 = -3.;

   Y1 = 3.;

   X2 = -7.;

   Y2 = 7.;

 

   area1 = \textbf{ellipse\_line\_overlap} (PHI, A, B, H, K, X1, Y1, X2, Y2, \&rtn);

   \textbf{sprintf} (msg,"Fig 4: area = \%15.8f, return_value = \%d\textbackslash n", area1, rtn);

   \textbf{printf} (msg);

   

   //-- case shown in Fig. 4, points reversed

   A = 4.;

   B = 2.;

   H = -6;

   K = 3;

   PHI = 3.*pi/8.0;

   X1 = -7.;

   Y1 = 7.;

   X2 = -3.;

   Y2 = 3.;

 

   area2 = \textbf{ellipse\_line\_overlap} (PHI, A, B, H, K, X1, Y1, X2, Y2, \&rtn);

   \textbf{sprintf} (msg,"Fig 4 reverse: area = %15.8f, return_value = \%d\textbackslash n", area2, rtn);

   \textbf{printf} (msg);

 

   \textbf{sprintf} (msg,"sum of ellipse segments = %15.8ftextbackslash n", area1 + area2);

   \textbf{printf} (msg);

   \textbf{sprintf} (msg,"total ellipse area by pi*a*b = %15.8ftextbackslash n", pi*A*B);

   \textbf{printf} (msg);

 

   return rtn; 

 }

 

 

 call_ee.c:

 

 #include $<$stdio.h$>$

 #include "program_constants.h"

 double ellipse_ellipse_overlap (double PHI_1, double A1, double B1, 

                                 double H1, double K1, double PHI_2, 

                                 double A2, double B2, double H2, double K2, 

                                 int *rtnCode);

                              

 int main (int argc, char ** argv) 

 {

    double A1, B1, H1, K1, PHI_1;

    double A2, B2, H2, K2, PHI_2;

    double area;

    int rtn;

    char msg[1024];

    printf ("Calling ellipse_ellipse_overlap.c\textbackslash n\textbackslash n");

    

    //-- case 0-1

    A1 = 3.; B1 = 2.; H1 = 0.; K1 = 0.; PHI_1 = 0.;

    A2 = 2.; B2 = 1.; H2 = -.75; K2 = 0.25; PHI_2 = pi/4.;

    area = ellipse_ellipse_overlap (PHI_1, A1, B1, H1, K1, 

                                    PHI_2, A2, B2, H2, K2, \&rtn);

    sprintf (msg,"Case 0-1: area = \%15.8f, return_value = \%d\textbackslash n", area, rtn);

    printf (msg);

    sprintf (msg,"          ellipse 2 area by pi*a2*b2 = \%15.8f\textbackslash n", pi*A2*B2);

    printf (msg);

    

    //-- case 0-2

    A1 = 2.; B1 = 1.; H1 = 0.; K1 = 0.; PHI_1 = 0.;

    A2 = 3.; B2 = 2.; H2 = -.3; K2 = -.25; PHI_2 = pi/4.;

    area = ellipse_ellipse_overlap (PHI_1, A1, B1, H1, K1, 

                                    PHI_2, A2, B2, H2, K2, &rtn);

    sprintf (msg,"Case 0-2: area = %15.8f, return\_value = \%d\textbackslash n", area, rtn);

    printf (msg);

    sprintf (msg,"          ellipse 1 area by pi*a1*b1 = \%15.8f\textbackslash n", pi*A1*B1);

    printf (msg);

    

    //-- case 0-3

    A1 = 2.; B1 = 1.; H1 = 0.; K1 = 0.; PHI_1 = 0.;

    A2 = 1.5; B2 = 0.75; H2 = -2.5; K2 = 1.5; PHI_2 = pi/4.;

    area = ellipse_ellipse_overlap (PHI_1, A1, B1, H1, K1, 

                                    PHI_2, A2, B2, H2, K2, &rtn);

    sprintf (msg,"Case 0-3: area = \%15.8f, return_value = \%d\textbackslash n", area, rtn);

    printf (msg);

    printf ("          Ellipses are disjoint, ovelap area = 0.0\textbackslash n\textbackslash n");

    

    //-- case 1-1

    A1 = 3.; B1 = 2.; H1 = 0.; K1 = 0.; PHI\_1 = 0.;

    A2 = 2.; B2 = 1.; H2 = -1.0245209260022; K2 = 0.25; PHI_2 = pi/4.;

    area = ellipse_ellipse_overlap (PHI_1, A1, B1, H1, K1, 

                                    PHI_2, A2, B2, H2, K2, \&rtn);

    sprintf (msg,"Case 1-1: area = \%15.8f, return\_value = \%d\textbackslash n", area, rtn);

    printf (msg);

    sprintf (msg,"          ellipse 2 area by pi*a2*b2 = \%15.8f\textbackslash n", pi*A2*B2);

    printf (msg);

    

    //-- case 1-2

    A1 = 2.; B1 = 1.; H1 = 0.; K1 = 0.; PHI_1 = 0.;

    A2 = 3.5; B2 = 1.8; H2 = .22; K2 = .1; PHI_2 = pi/4.;

    area = ellipse_ellipse_overlap (PHI_1, A1, B1, H1, K1, 

                                    PHI_2, A2, B2, H2, K2, \&rtn);

    sprintf (msg,"Case 1-2: area = \%15.8f, return_value = \%d\textbackslash n", area, rtn);

    printf (msg);

    sprintf (msg,"          ellipse 1 area by pi*a1b1 = \%15.8f\textbackslash n", pi*A1*B1);

    printf (msg);

    

    //-- case 1-3

    A1 = 2.; B1 = 1.; H1 = 0.; K1 = 0.; PHI_1 = 0.;

    A2 = 1.5; B2 = 0.75; H2 = -2.01796398085; K2 = 1.25; PHI_2 = pi/4.;

    area = ellipse_ellipse_overlap (PHI_1, A1, B1, H1, K1, 

                                    PHI_2, A2, B2, H2, K2, \&rtn);

    sprintf (msg,"Case 1-3: area = %15.8f, return\_value = \%d\textbackslash n", area, rtn);

    printf (msg);

    printf ("          Ellipses are disjoint, ovelap area = 0.0\textbackslash n\textbackslash n");

    

    //-- case 2-1

    A1 = 3.; B1 = 2.; H1 = 0.; K1 = 0.; PHI_1 = 0.;

    A2 = 2.25; B2 = 1.5; H2 = 0.; K2 = 0.; PHI_2 = pi/4.;

    area = ellipse_ellipse_overlap (PHI_1, A1, B1, H1, K1, 

                                    PHI_2, A2, B2, H2, K2, \&rtn);

    sprintf (msg,"Case 2-1: area = \%15.8f, return_value = \%d\textbackslash n", area, rtn);

    printf (msg);

    sprintf (msg,"          ellipse 2 area by pi*a2*b2 = \%15.8f\textbackslash n", pi*A2*B2);

    printf (msg);

    

    //-- case 2-2

    A1 = 2.; B1 = 1.; H1 = 0.; K1 = 0.; PHI_1 = 0.;

    A2 = 3.; B2 = 1.7; H2 = 0.; K2 = 0.; PHI_2 = pi/4.;

    area = ellipse_ellipse_overlap (PHI_1, A1, B1, H1, K1, 

                                    PHI_2, A2, B2, H2, K2, \&rtn);

    sprintf (msg,"Case 2-2: area = \%15.8f, return_value = \%d\textbackslash n", area, rtn);

    printf (msg);

    sprintf (msg,"          ellipse 1 area by pi*a1b1 = \%15.8f\textbackslash n", pi*A1*B1);

    printf (msg);

    

    //-- case 2-3

    A1 = 3.; B1 = 2.; H1 = 0.; K1 = 0.; PHI_1 = 0.;

    A2 = 2.; B2 = 1.; H2 = -2.; K2 = -1.; PHI_2 = pi/4.;

    area = ellipse_ellipse_overlap (PHI_1, A1, B1, H1, K1, 

                                    PHI_2, A2, B2, H2, K2, \&rtn);

    sprintf (msg,"Case 2-3: area = \%15.8f, return\_value = \%d\textbackslash n\textbackslash n", area, rtn);

    printf (msg);

    

    //-- case 3-1

    A1 = 3.; B1 = 2.; H1 = 0.; K1 = 0.; PHI_1 = 0.;

    A2 = 3.; B2 = 1.; H2 = 1.; K2 = 0.35; PHI_2 = pi/4.;

    area = ellipse_ellipse_overlap (PHI_1, A1, B1, H1, K1, 

                                    PHI_2, A2, B2, H2, K2, \&rtn);

    sprintf (msg,"Case 3-1: area = \%15.8f, return\_value = \%d\textbackslash n", area, rtn);

    printf (msg);

    

    //-- case 3-2

    A1 = 2.; B1 = 1.; H1 = 0.; K1 = 0.; PHI_1 = 0.;

    A2 = 2.25; B2 = 1.5; H2 = 0.3; K2 = 0.; PHI_2 = pi/4.;

    area = ellipse_ellipse_overlap (PHI_1, A1, B1, H1, K1, 

                                    PHI_2, A2, B2, H2, K2, \&rtn);

    sprintf (msg,"Case 3-2: area = \%15.8f, return\_value = \%d\textbackslash n\textbackslash n", area, rtn);

    printf (msg);

    

    //-- case 4-1

    A1 = 3.; B1 = 2.; H1 = 0.; K1 = 0.; PHI_1 = 0.;

    A2 = 3.; B2 = 1.; H2 = 1.; K2 = -0.5; PHI_2 = pi/4.;

    area = ellipse_ellipse_overlap (PHI_1, A1, B1, H1, K1, 

                                    PHI_2, A2, B2, H2, K2, \&rtn);

    sprintf (msg,"Case 4-1: area = \%15.8f, return_value = \%d\textbackslash n", area, rtn);

    printf (msg);

    

    return rtn; 

 }
\end{lstlisting}


\medskip

\end{document}